\documentstyle[12pt]{article}

\newcommand{\resection}[1]{\setcounter{equation}{0}\section{#1}}

\def\Z {\bf Z}
\def\c {\tilde{\chi}}
\def\s {\sigma}
\def\q {\bar{q}}
\def\e {\epsilon}
\def\be{\begin{equation}}
\def\ee{\end{equation}}
\def\ba{\begin{array}}
\def\ea{\end{array}}
\def\EQ{\begin{equation}}
\def\EN{\end{equation}}
\def\bea{\begin{eqnarray}}
\def\eea{\end{eqnarray}}
\def\to{\rightarrow}

\def\sa{\hspace{0.1in}}

\def\sb{\hspace{0.2in}}
\def\sc{\hspace{0.3in}}

\def\O{{\cal O}}
\def\Q{{\cal Q}}
\def\F{{\cal F}}
\def\K{{\cal K}}
\def\M{{\cal M}}
\def\P{{\cal P}}

\begin{document}
\oddsidemargin 5mm
\setcounter{page}{0}
\renewcommand{\thefootnote}{\fnsymbol{footnote}}
\newpage
\setcounter{page}{0}
\begin{titlepage}
\begin{flushright}
DAMTP-HEP-94/85
\end{flushright}
\vspace{0.5cm}
\begin{center}
{\large {\bf The Space of Local Operators in Perturbed Conformal Field
Theories}}\\
\vspace{1.5cm}
{\bf
A. Koubek\footnote{E-mail: a.koubek@amtp.cam.ac.uk} }\\
\vspace{0.8cm}
{\em Department of Applied Mathematics and Theoretical Physics,\\
Silver Street,\\
CB3 9EW Cambridge, \\UK} \\
\end{center}
\vspace{6mm}
\begin{abstract}
  The space of local operators in massive deformations of conformal
  field theories is analysed. For several model systems it is shown
  that one can define chiral sectors in the theory, such that the
  chiral field content is in a one-to-one correspondence with that of
  the underlying conformal field theory.  The full space of operators
  consists of the descendent spaces of all scalar fields. If the
  theory contains asymptotic states which satisfy generalised
  statistics, the form factor equations admit in addition also
  solutions corresponding to the descendent spaces of the
  para-fermionic operators of the same spin as the asymptotic states.
  The derivation of these results uses $q$-sum expressions for the
  characters and $q$-difference equations used in proving
  Rogers-Ramanujan type identities.
\end{abstract}
\vspace{5mm}
\begin{center}
October 1994
\end{center}
\end{titlepage}
\newpage

\resection{Introduction}

During the last several years the theory of completely integrable
models in $(1+1)$ dimensions has experienced a  rapid
development.  Especially the study of {\em conformal field theories}
(CFT), which resulted in the solution of several classes of such
theories, has allowed a deep understanding of the critical points
describing second order phase transitions.  Unfortunately, less is
known for off-critical integrable field theories.  Their solution
should give a description of the scaling region around critical points
and allow a deeper understanding of standard techniques used in
quantum field theory, such as the renormalisation group or
perturbation theory.

In this article we will study so-called {\em perturbed conformal field
  theories}, {\em i.e.} models which can be viewed as perturbations of
a conformal field theory by a relevant operator. The perturbation is
chosen in a way that the theory remains integrable also away from the
critical point. This description proves useful, since many properties
of the conformal theory can be preserved or 'deformed'
off-criticality. As, for example, the quantum group structure, which
becomes an affine one in the perturbation process, or the conserved
charges.

The ultimate goal in the investigation of these systems is to solve
them, that is to determine the algebra of local operators and their
correlation functions. In general conformal field theories perturbed
by the operators $\phi_{1,2}$, $\phi_{2,1}$ and $\phi_{1,3}$ \be
H=H_{CFT} + \lambda \int d^2 x\, \phi (x) \label{hpert} \ee are driven
into a massive regime.  It has been shown by A. Zamolodchikov
\cite{Zam}, that such theories can be described by the exact
scattering matrix. A lot of progress has been made in determining in
general the on-shell structure and understanding its algebraic origin.

The understanding of the off-shell structure of these massive
integrable theories is less complete. Knowing the exact $S$-matrix of
a theory one can calculate the matrix-elements of local operators,
called form factors. They are determined from the knowledge of the
on-shell structure of the theory, through the {\em form factor
  equations}.  This is a remarkable property, since the on-shell
structure of the theory does not contain any information about local
fields. In fact it is described in terms of asymptotic states.

The form factor equations do not refer to a specific operator. This on
the one hand makes the identification of operators a difficult task.
Apart from the fundamental fields of a theory (as the energy-momentum
tensor or conserved currents) which have additional properties, it is
difficult to identify a specific operator from the knowledge of its
form factors. This is true especially in this case, where we are
interested in the comparison with the conformal field theory in the
ultraviolet limit. In order to achieve such an identification, one would
need to calculate correlation functions, which involves the summing up
of the form factor series, a problem which has not been solved for
interacting models.

The above mentioned arbitrariness in the form factor equations allows
the determination of the space of local operators of the theory. Since
the equations must be valid for {\em all} local fields of the theory,
one can specify the field content by determining the independent
solutions of the form factor equations.  In this article we will use
precisely this fact, and find the space of local operators for some
simple massive theories. They have the distinguishing property that
they contain only scalar asymptotic states, and can be described as
perturbed minimal conformal field theories.

Beforehand, it is not clear what kind of structure one should expect
to find in the perturbed conformal theories. Since the perturbation
theory defined by (\ref{hpert}) is super-renormalizable, only a finite
number of counter terms is involved in the renormalisation procedure,
which should leave unchanged the structure of the operator algebra of
the theory \cite{Zam}. On the other hand the most salient feature of
the critical theory is its Virasoro algebra symmetry, which is broken
in the massive theory and one expects drastic changes in the
off-critical theory.  In the critical theory left and right chiral
sectors (corresponding to the components $T$ and $\bar{T}$ of the
energy momentum tensor) are independent which certainly is not true in
the deformed theory. Also the structure of the conserved charges
changes in the course of the perturbation.

The result of this study will be that the structure of the massive
theory resembles much the one of the conformal theory. We will define
a grading in the space of operators which allows a formal definition
of chiral components.  The so-defined chiral structure of the deformed
theory becomes isomorphic to the conformal one.  In CFT on the real
line there are no constraints on which combinations of the two chiral
dimensions of the operators can enter the partition function, apart
that they should form a closed operator algebra. In the massive model
we have an additional locality requirement, deriving from the form
factor equations. As will be shown in the following sections, only
specific combinations of chiral dimensions are present in the massive
theory.

\subsection{The Form Factor Bootstrap Approach}

Let us review some general properties of the form factors
\cite{Karowski,nankai,Yurov-Zam,berle}.  We parametrise the momenta in
terms of the masses of the particles $m_i$ and the rapidity variables
$\beta_i$ as
\[
p_i^0\,=\,m_i\cosh\beta_i\,\,\, , \,\,\,
p_i^1\,=\,m_i\sinh\beta_i\,\,\, .
\]
Form factors are defined as matrix elements of local operators $\O$
between the vacuum and the set of asymptotic states, \EQ
\F_{\e_1,\e_2,\dots,\e_n}^{\cal O} (\beta_1,\beta_2,\ldots,\beta_n)
\,=\, \langle 0\mid{\cal O}(0,0)\mid Z_{\e_1}( \beta_1),
Z_{\e_2}(\beta_2),\ldots,Z_{\e_n}(\beta_n)\rangle_{in}\sb .
\label{FoF}
\EN Properties deriving from basic principles of quantum field theory,
such as crossing and unitarity, and the properties of the
Fadeev-Zamolodchikov operators $Z_{\e}(\beta)$ allow the construction
of these matrix elements.  For the case of scalar particles these
functional equations known as the {\em form factor axioms} are:\\

\noindent 1) Watson's equations \bea & &
\F^\O_{\e_1\dots\e_i\e_{i+1}\dots\e_n}
(\beta_1,\dots,\beta_i,\beta_{i+1},\dots\beta_n) =\nonumber \\ &
&S_{\e_i\e_{i+1}}(\beta_i-\beta_{i+1})\,
\F^\O_{\e_1\dots\e_{i+1}\e_{i}\dots\e_n}
(\beta_1,\dots,\beta_{i+1},\beta_{i},\dots\beta_n) \sb ,
\label{wat1}\eea

\be \F^\O_{\e_1\e_2\dots\e_n} (\beta_1+2\pi i,\beta_2,\dots,\beta_n )=
e^{2 i \pi (s_{\e_n}+w_{\e_n} +\sum w_{\e_i\e_n})} \F^\O_{\e_2\dots\e_n\e_1}
(\beta_2,\dots,\beta_n,\beta_1 ) \sb ,
\label{wat2} \ee
where $s_{\e_n}$ denotes the spin of the particle $\e_n$ and $w_\e$
denotes the mutual locality index between the field $\O$ and the field
creating the asymptotic state $Z_{\e}$.  Similarly $w_{\e_i\e_n}$
denotes the generalised statistics, which the operators creating the
particles $i$ and $n$ satisfy. It is determined by $\lim_{\beta \to
  \infty} S_{ni} (\beta)$. This phase factor is necessary since for
some theories we will consider, the particles will not become free
particles for $\beta \to \pm \infty$, but will satisfy
$$Z_{\e_i} Z_{\e_j} = e^{2 i \pi w_{ij}} Z_{\e_j} Z_{\e_i} \sb .$$
Therefore whenever one continues the form factors analytically, one
needs to take these commutation properties into account.  For theories
containing degenerate particles this statistics is governed by the
$R$-matrix defining the ultraviolet limit of the scattering theory
\cite{berle}.  Alternatively these phases can be taken in account by
assigning non-trivial charge conjugation to the operators creating the
asymptotic states.

\noindent 2) Relativistic invariance \be \F^\O_{\e_1\dots\e_n}
(\beta_1+\Lambda,\dots,\beta_n+\Lambda ) = e^{s\Lambda}
\F^\O_{\e_1\dots\e_n} (\beta_1,\dots,\beta_n ) \sb .\label{prop2}\ee

\noindent 3) Asymptotic behaviour \be
\F_{\e_1,\dots,\e_n}(\beta_1+\Lambda,\dots,\beta_i +\Lambda,
\beta_{i+1}, \dots , \beta_{n}) = O(e^{S_n^i |\Lambda |}) \sb {\rm for}
\sb |\Lambda| \sim \infty \sb ,
\label{asym}\ee
and $S \ \equiv \max(S_n^i)$ is a finite constant.

\noindent
4) Equation of kinematical poles
$$ -i \lim_{\beta'\to \beta}(\beta'-\beta) \F^\O_{\e\e\e_1\dots\e_n}
(\beta'+i\pi,\beta,\beta_1,\dots,\beta_n) =$$ \nopagebreak \be \left
(1-e^{2 i \pi (w_\e+\sum w_{\e_i\e})} \prod_{i=1}^{n}
S_{\e\e_i}(\beta-\beta_i) \right ) \F^\O_{\e_1\dots\e_n}
(\beta_1,\dots,\beta_n ) \sb .\label{kin} \ee

\noindent 5) Bound state equation: Let particles $A_i$, $A_j$ form a
bound state $A_k$, the corresponding two-particle scattering amplitude
exhibits a pole with the residue \be -i \lim_{\beta'\to
  iu_{\e_i\e_j}^{\e_k}}(\beta-iu_{\e_i\e_j}^{\e_k})
S_{\e_i\e_j}(\beta) =(\Gamma_{\e_i\e_j}^{\e_k})^2 \sb ;
\label{residueS}\ee
then
$$ -i \lim_{\beta'\to \beta}(\beta'-\beta)
\F^\O_{\e_1\dots\e_i\e_j\dots\e_n} (\beta_1,\dots,\beta'+i\bar
u_{\e_i\e_k}^{\e_j},\beta-i\bar u_{\e_j\e_k}^
{\e_i},\dots,\beta_{n-1}) = $$ \be =\Gamma_{\e_i\e_j}^{\e_k}
\F^\O_{\e_1\dots\e_k\dots\e_n} (\beta_1,\dots,\beta,\dots,\beta_{n-1}
) \sb . \label{bounds}\ee

In \cite{Smirnov1} it was shown that the operators defined by the form
factors (\ref{FoF}) satisfy proper locality relations provided that
they satisfy the properties (\ref{wat1})-(\ref{asym}) and the residue
equations (\ref{kin}) and (\ref{bounds}).

A well established solution method uses the parametrisation of the
form factors in terms of the {\em minimal two particle form factor}
and the explicit prescription of the pole structure.  As a first step
one determines the minimal two particle form factor, which satisfies
Watson's equations for $n=2$, \be \F^{\rm min}_{ab} (\beta_{12}) =
S_{ab} (\beta_{12}) \F^{\rm min}_{ba} (-\beta_{12}) \sa,\sb \F^{\rm
  min}_{ab}(i\pi - \beta_{12}) = \F^{\rm min}_{ba}(i \pi+\beta_{12})
\sb ,\label{w2}\ee where $\beta_{12} =\beta_1-\beta_2$. In order to
specify an unique solution to these equations one imposes further
restrictions, namely that $F^{\rm min}$ is analytic in $0<{\rm Im}
\beta < \pi$ and has no zeros in this range.  The $n$--particle form
factor can then be parametrised as
$$ \F_{\e_1,\dots,\e_n} (\beta_1,\dots,\beta_n) = $$ \be
H_{\e_1,\dots,\e_n} Q_{\e_1,\dots,\e_n}(\beta_1,\dots,\beta_n)
\prod_{i<j} \frac{\F^{\rm min}_{\e_i\e_j}(\beta_{\e_i\e_j})}{
  (x_i+x_j)^{\delta_{\e_i \e_j}}
\prod_{\alpha} \left ( \sinh \frac 12 ( \beta_{ij} - i
  \alpha ) \sinh \frac 12 ( \beta_{ij} + i \alpha )\right )} \sb
.\label{paraf}\ee The product over $\alpha$ runs over all bound state
poles deriving from the $S$--matrix; $H_{\e_1,\dots,\e_n}$ are
normalisation constants, and $x_i \equiv e^{\beta_i}$.

The only unknowns are therefore the functions
$Q_{\e_1,\dots,\e_n}(x_1,\dots,x_n)$.  They are homogenous functions,
analytic apart from the origin and polynomially bounded because of
(\ref{asym}). They are symmetric functions in their arguments $x_i$
and $x_j$ if $\e_i=\e_j$.

Let us focus our attention on the fundamental particle of the theory.
In sections \ref{sec-ising}-\ref{sec-35} we will deal with systems
containing only one asymptotic state and therefore only form factors
with all indices corresponding to this particle have to be considered.
Denote the corresponding function $Q_{11\dots1} = Q_n$, for $n$
particles. This is, according to the above discussion, a totally
symmetric function.  Therefore we introduce the elementary symmetric
polynomials, which form a base in the space of symmetric polynomials
and are generated by \cite{Macdon} \EQ \prod_{i=1}^n(x+x_i)\,=\,
\sum_{k=0}^n x^{n-k} \,\sigma_k(x_1,x_2,\ldots,x_n) \sb .
\label{generating} \EN Further we will use the notation that $\bar{x}
= 1/x$ and $\bar{\s}_i(x_1,\dots ,x_n)  =\s_i(\bar{x}_1,\dots\bar{x}_n)$.
Then the functions $Q_n$ can be cast into the following form \cite{CM}
\be Q_n(x_1,\dots,x_n) = \sum_N \frac{1}{\s_n^{N-\gamma}} P_n (x_1,
\dots,x_n) \sb ,
\label{ansatz} \ee wherein $\gamma$ denotes a possible non-integer
power which is determined from the spin of the corresponding operator
through (\ref{wat2}) and (\ref{prop2}).  The polynomials $P_n$ can be
expressed as polynomials in the elementary symmetric polynomials
$\s_i$.

Before turning to the explicit construction of solutions of the form
factor equations for specific systems, we want to discuss the general
structure of the space of local operators.  Suppose we have calculated
the form factors of a scalar operator (for example the trace of the
energy momentum tensor). Representing the operator in terms of its
form factors one finds that the form factors of the derivative
operators $\partial_z \O$ ( $\partial_{\bar{z}} \O$) are given by $\s_1
\F_n$ ($\bar{\s}_1 \F_n$).  This indicates, that in the massive model
the two chiral sectors are represented by the variables $x_i$ and
$\bar{x}_i$ respectively.  But since \be\bar{\s}_i(x_1,\dots,x_n) =
\frac{\s_{n-i}(x_1,\dots,x_n)}{\s_n (x_1,\dots,x_n)}\sb ,\label{sbar}
\ee any $\bar{x}$ dependence can be rewritten in terms of $x$,
reflecting the fact that right and left chiral sectors are {\em not}
independent in the massive theory.

We want to introduce a grading into the space of operators that
resembles the structure of highest weight representations of the
Virasoro algebra, in order to find their massive counter-parts.
It can be seen from (\ref{paraf}), that increasing the power $N$ in
(\ref{ansatz}) for fixed spin $s$, will increase the divergence of the
corresponding form factor $\F_n$.  For that we define as the chiral
(left) operators those with the mildest ultraviolet behaviour, which
means we put $N=0$ in (\ref{ansatz}).  Similarly we conjecture that
augmenting $N$ increases the presence of right components in the
derivative state. A simple example of this fact is $\bar{L}_{-1} \phi
= \bar{\partial}\phi$ as we discussed above.  Unfortunately we cannot
assign a one-to-one correspondence between conformal states $(h,\bar{
  h})$ and massive states $(N+s,N)$. The parameter $N$ rather
indicates how many operators $\bar{L}$ act on the primary operator,
but also depends on the explicit representation of $\bar{L}_n$ in the
massive theory. We will discuss this fact in detail for specific
examples.

Finally we want to stress that setting the scale N is a
rather subtle issue. Because in every module of chiral massive
descendent operators, there will be polynomials $P_n$ which are
proportional to $\s_n$. Therefore one could identify those as chiral
descendents of the primary operator, and regard the original ones as
corresponding to $N=1$ states. For some operators this ambiguity can
be solved, since one knows the form factors of the primary operators
explicitely. But as we will see in section \ref{sec-35}, this scale is
not necessarily the same for all primary operators in a theory.

In the next section we analyse the scaling field theory near the
critical point of the Ising model, in zero magnetic field, $\M_{3,4}+
\phi_{1,3}$.  Because of its simple structure (it can be described as
a massive free fermion theory), we can introduce our method and test
it against known results.  In section \ref{sec-yl} we investigate the
Yang-Lee model, as the simplest example of an interacting field
theory.  In section \ref{sec-35} we analyse another non-unitary one
particle model, $\M_{3,5}+\phi_{1,3}$.  Finally, in section
\ref{sec-2n} we generalise these results to a series of models, namely
$\M_{2,2n+3} +\phi_{1,3}$.  As we will see the physical picture
persists, while the mathematical methods become more involved.

\resection{ The Ising Model\label{sec-ising}}

As a first step we will apply the form factor bootstrap to the scaling
field theory near the critical point of the two dimensional Ising
model in zero magnetic field.  This massive field theory can be
described in terms of free Majorana fermions, or equally in terms of a
bosonic particle which interacts through an $S$-matrix, $S=-1$. The
theory is divided into two sectors. One contains the monomials in the
fermion fields, and correlation functions of such operators can be
calculated by using Wick's theorem. The other sector contains fields
which aquire a phase $e^{i \pi}$ when moved around the fermions. The
basic fields in this sector are the order and disorder fields $\s (x)
$ and $\mu(x)$.  Their correlation functions have been analysed in
numerous ways, and using the form factor approach in
\cite{Yurov-Zam,CM}.

For the scaling Ising model the parametrisation (\ref{paraf}) takes
the explicit form \cite{Yurov-Zam,CM,MTS} \be
\F_n(\beta_1\dots\beta_n) = Q_n(x_1,\dots,x_n) H_n (\s_n)^{\frac 12
  \delta_{w,0}}\frac 1{(\s_n)^N} \prod_{i<j}^n \tanh \frac
{(\beta_i-\beta_j)}{ 2}
\label{pi}\sb . \ee
This parametrisation reduces the equation of kinematical poles
(\ref{kin}) to \be\ba{ll} Q_{n+2} (-x,x,x_1,\dots,x_n) = x^{2 N} Q_n
(x_1,\dots,x_n)\sa , & w=\frac 12\sb ,
\label{recisingo} \\
Q_{n+2} (-x,x,x_1,\dots,x_n) = 0 \sa , & w= 0\sb . \ea \ee Notice the
half-integer power of $\s_n$ which appears in the sector $w=0$.  It
has to be introduced in order to satisfy (\ref{wat2}), and has the
effect that form factors with odd index in this sector will carry half
integer spin.

The sector $w=\frac 12$ contains the order and the disorder fields
$\s$ and $\mu$.  The form factors of both operators have been
calculated and are given by $Q_n=const.$ in (\ref{pi}).  In \cite{CM}
the descendent operators of $\s(x)$ have been determined. They were
constructed by building symmetric polynomials of spin $s$, which are
invariant under the recursion relation (\ref{recisingo}). Only
invariants with odd spin are consistent with the recursion relation,
and can be expressed in terms of the elementary symmetric polynomials
as the $n+1$ dimensional determinant \cite{Christe} \be I_{2 n +1} =
\left \vert \ba{ccccc} \s_1 & \s_3 & \s_5 & \s_7 & \dots
\\ 1 & \s_2 & \s_4 & \s_6 & \dots \\ 0 & 1 & \s_2 & \s_4 & \dots \\ 0
& 0 & 1 & \s_2 & \dots \\ \vdots & \vdots & \vdots & \vdots & \ddots
\ea \right \vert\sb .\label{cons} \ee Arbitrary polynomials formed
from the quantities (\ref{cons}) constitute solutions to
(\ref{recisingo}), which is the equivalent situation as in conformal
field theory where the Virasoro irreducible representation of the
operator $\phi_{1,2}$ is generated by the application of the operators
$L_{-(2 n+1)}$ to the primary field.

The method we want to describe chooses a different basis in the single
subspaces of spin $s$. Namely we use the fact that the number of
solutions of the recursion relations at level $n$ is given by the
number of solutions at level $n-2$ plus the dimension of the kernel of
the recursion relations (\ref{recisingo}).  This amounts in a simple
counting procedure which only involves the comparison of the degree of
the polynomials $Q_n$ and that of the kernel\footnote{Note that the
  second line in (\ref{recisingo}) is precisely the definition of the
  kernel of the recursion relation.}. This has the advantage that one
finds {\em all} solutions to the form factor equations, and not only
those which can be considered as descendent operators of some
non-trivial primary field. In the case of the Ising model the kernel
of the recursion relation (\ref{recisingo}) is given by \be \K_n =
\prod_{i<j}^n (x_i +x_j) \sb .\label{kernel}\ee The total degree of
this function is $deg(\K_n )= \frac 12 n (n-1)$.

Comparing the degree of the polynomial $Q_n$ and the degree of $\K_n$,
we can now construct explicitly solutions of the recursion relations.
For the odd form factors, we find the following solutions for the
first few values of spin $s$: \be\ba{cl} s=0 &Q_1= const. \sb ,\\ s=1
&Q_1= \s_1 \sb ,\\ s=2 &Q_1= \s_1^2 \sb ,\\ s=3 &Q_1= \s_1^3$, $Q_3 =
\K_3 \sb , \ea\label{count1}\ee which indicates that we have $1,1,1,2$
linear independent solutions at spin levels $s=0,1,2,3$ respectively.
Notice that these values correspond exactly to the first terms of the
character expansion of the conformal field $\phi_{1,2}$, namely
$\c_{1,2} = 1+q+q^2+2 q^3\dots$.

We introduce the character $\chi$ of the conformal minimal model as
\be \chi_{r,s}^{(p,q)} = q^{(h-c/24)} \c_{r,s}^{(p,q)} \sb , \ee where
$c$ and $h$ denote the central charge and the conformal dimension of
the primary operator respectively. Therefore $\c$ encodes the
degeneracies in the Virasoro representation \cite{rocha}, \be
\c_{r,s}^{p,q} = \frac1{(q)_\infty}\sum_{\mu=-\infty}^\infty (q^{\mu
  (\mu p q + r q - s p)}- q^{(\mu p+r)(\mu q+s)})\sb .
\label{chardef} \ee
Further, define $\P(m,n)$ to be the number of partitions of $n$ into
numbers whose value do not exceed $m$. These partitions are generated
by \be \frac 1{(q)_m} = \sum \P(m,n) q^n \label{parts}\sb ,\ee where
\be(q)_m \equiv \prod_{i=1}^m (1-q^i) \sb .\label{GF}\ee

Using the definition (\ref{parts}), we rewrite our counting of the
solutions in (\ref{count1}) in the following way. At $s=0$ the number
of solutions is given by $\P(1,0)$, for $s=1$ by $\P(1,1)$, for $s=2$
by $\P(1,2)$ and for $s=3$ by $\P(1,3)+\P(3,0)$. Using this notation
and the generating function (\ref{GF}), one immediately finds the
number of solutions at general spin $s$, as being generated by \be
F_0\equiv\sum_{m,\, odd} \frac{q^{\frac 12 m (m-1)}}{(q)_m} =
\c_{1,2}^{(3,4)}(q) \sb .\ee Note that this formalism leads rather to
`fermionic' expressions of the character than to the `bosonic' form
(\ref{chardef}). This notation ( `fermionic' and `bosonic' ) has been
introduced by R. Kedem {\em et.al.} \cite{kedem} in their study of
character identities for a large class of conformal field theories.
Only recently bases have been constructed for conformal field theories
which lead to fermionic sum expressions for the characters \cite{sh}.
It is an interesting fact that the form factor approach leads to them
in a straightforward way.

Similarly one can analyse the even form factors, being related to the
disorder field $\mu$. The counting procedure can be carried out
analogously, and one finds that the number of chiral operators for
spin $s$ is generated by \be G_0\equiv\sum_{m,\, even} \frac{q^{\frac
    12 m (m-1)}}{(q)_m} = \c_{1,2}^{(3,4)}(q) \sb .\ee As it is well
known, both order and disorder fields are related to the conformal
operator $\phi_{1,2}$.

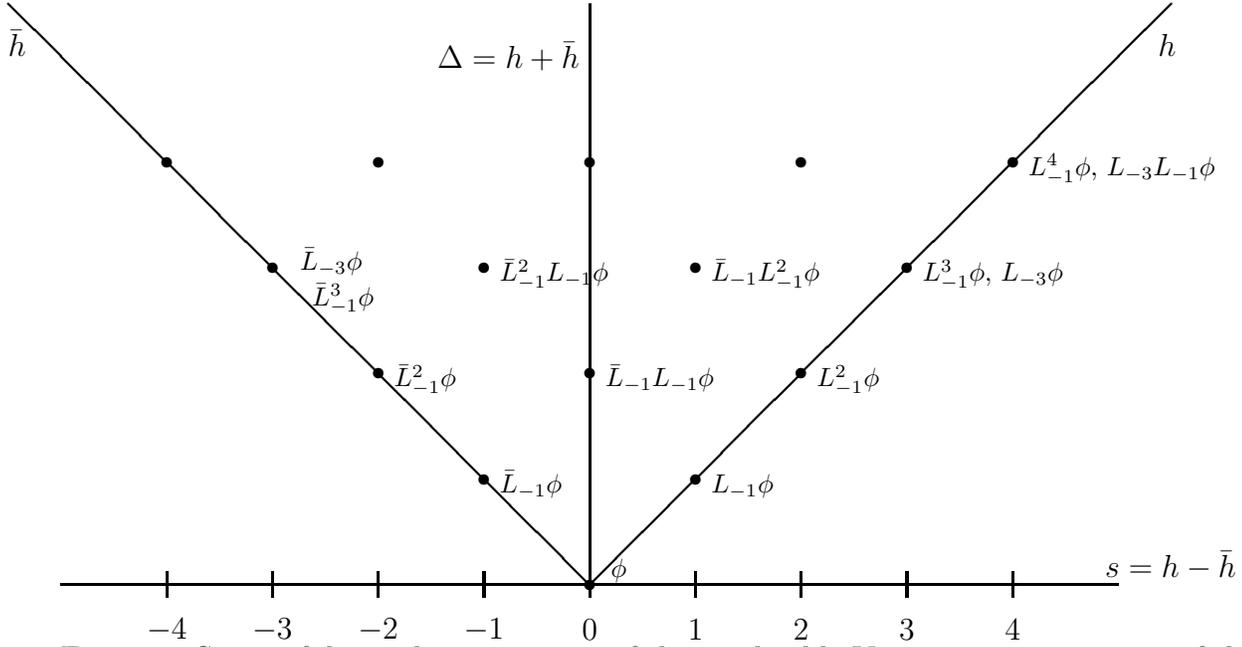
\begin{figure}
\begin{picture}(400,250)(-200,0)
\thicklines
\put(-200,0){\line(1,0){400}}
\put(0,0){\line(0,1){220}}
\put(0,0){\line(1,1){220}}
\put(0,0){\line(-1,1){220}}
\multiput(0,0)(40,40){5}{\circle*{4}}
\multiput(-40,40)(40,40){4}{\circle*{4}}
\multiput(-80,80)(40,40){3}{\circle*{4}}
\multiput(-120,120)(40,40){2}{\circle*{4}}
\multiput(-160,160)(40,40){1}{\circle*{4}}
\put(215,200){$h$}
\put(-220,200){$\bar{h}$}
\put(-4,200){\makebox(0,0)[r]{$\Delta = h + \bar{h}$}}
\multiput(-160,-5)(40,0){9}{\line(0,1){10}}
\put(-160,-17){\makebox(0,0){$-4$}}
\put(-120,-17){\makebox(0,0){$-3$}}
\put(-80,-17){\makebox(0,0){$-2$}}
\put(-40,-17){\makebox(0,0){$-1$}}
\put(0,-17){\makebox(0,0){$0$}}
\put(40,-17){\makebox(0,0){$1$}}
\put(80,-17){\makebox(0,0){$2$}}
\put(120,-17){\makebox(0,0){$3$}}
\put(160,-17){\makebox(0,0){$4$}}
\put(195,3){$s=h-\bar{h}$}
\put(8,3){\footnotesize $\phi$}
\put(46,35){\footnotesize $L_{-1} \phi$}
\put(86,75){\footnotesize $L_{-1}^2 \phi$}
\put(126,115){\footnotesize $L_{-1}^3 \phi$, $L_{-3} \phi$}
\put(166,155){\footnotesize $L_{-1}^4 \phi$, $L_{-3} L_{-1}\phi$}
\put(-34,35){\footnotesize $\bar{L}_{-1} \phi$}
\put(6,75){\footnotesize $\bar{L}_{-1} L_{-1} \phi$}
\put(46,115){\footnotesize $\bar{L}_{-1} L_{-1}^2 \phi$}
\put(-74,75){\footnotesize $\bar{L}_{-1}^2 \phi$}
\put(-34,115){\footnotesize $\bar{L}_{-1}^2 L_{-1}\phi$}
\put(-114,115){\makebox(0,0)[l]{\footnotesize\shortstack{
 $\bar{L}_{-3} \phi$\\ $\sa\,\bar{L}_{-1}^3 \phi$}}}
\end{picture}
\caption{Space of descendent operators of the irreducible Virasoro
  representation of the field
  $\phi_{1,2}$ of the conformal Ising model.}
\label{fig-ising}
\end{figure}

As a next step we investigate solutions corresponding to $N=1$ in
(\ref{pi}). We conjecture that these operators correspond in the
ultraviolet limit to states of the type $\bar{L}_{-1} L_{-n_1} \dots
L_{-n_k} \vert \phi_{1,2} \rangle$ (compare figure \ref{fig-ising}).
Let us repeat the counting as we did for the primary operators. The
total degree of the polynomials $Q_n$ in this case is given by
$deg(Q_n)= n +s$ which must be compared with the degree of the kernel
of the recursion relations.

We list the first few solutions, \be \ba{ll} s=-1 & Q_1 =const. \sb
,\\ s=0 & Q_1 = \s_1 ,\, Q_3=\K_3 \sb ,\\ s=1 & Q_1 = \s_1^2 ,\,
Q_3=\s_1 \K_3 \sb ,\\ s=2 & Q_1 = \s_1^3 ,\, Q_3=\s_1^2 \K_3,\,
Q_3=\s_2 \K_3 \sb .  \ea \ee The counting here gives not only the
derivative states of the type $\bar{L}_{-1} L_{n_1} \dots L_{n_k}
\vert \phi_{1,2} \rangle$ as expected, but also the original chiral
ones.  For example from (\ref{pi}) one immediately realises that the
$N=1$ spin zero solution $Q_1 = \s_1$ coincides with the $N=0$ spin
zero solution $Q_1 = const.$, which is the one generating the form
factors of the primary operator.  In the same way every solution $Q_n$
being proportional to $\s_n$ corresponds to an element of the set of
chiral descendents.

The generating function of the $N=1$ states can be found in an
analogous way as before and is given by
$$ F_1 \equiv \sum_{m,odd}^{\infty}\frac{q^{\frac 12 m
    (m-3)}}{(q)_m}\sb . $$ In order to find the level one descendent
operators we need to subtract the chiral ones.
$$F_1-F_0 =   \sum_{m,odd}^{\infty}\frac{q^{\frac 12 m
    (m-3)}}{(q)_m} - \sum_{m,\, odd} \frac{q^{\frac 12 m (m-1)}}{(q)_m}$$
$$ \sum_{m,odd}^{\infty}\frac{q^{\frac 12 m (m-3)}(1-q^m)}{(q)_m} =
\sum_{m,odd}^{\infty}\frac{q^{\frac 12 m (m-3)}}{(q)_{m-1}} =
\sum_{m,even}^{\infty}\frac{q^{\frac 12 (m+1) (m-2)}}{(q)_{m}} = \q
G_0= \q \c_{1,2}(q)\sb ,$$ where we introduced the notation
$\q=q^{-1}$.  We have carried out the calculation explicitly, since
this re-summation technique is a fundamental tool in the analysis of
$q$-sum expressions and will be repeatedly used in the following. We
find that in the perturbed model we have the same number of
$\bar{h}=1+\frac 1{16}$ states as in conformal field theory. The
analysis of the $N=1$ states in the even sector is in complete
analogy, and will therefore be omitted.

Finally let us discuss operators in the subspace $N=2$, again for the
odd sector. Generalising from the analysis of the subspace $N=1$ one
might expect that in the ultraviolet limit the operators for $N=2$
correspond to states with $\bar{h}= 2 +\frac 1{16}$.  But for spin
$s=-2$ we find two solutions, which we denote by \be Q_1^{(1)} =
const.  \sa ,\sc Q_3^{(2)} = \s_1 \K_3 \sb ,\label{sol2}\ee whereas
from conformal field theory we would expect for $\bar{h}=2+\frac
1{16}$, $s=-2$ only {\em one} operator (compare figure
\ref{fig-ising}).  We conclude that the simple interpretation of $N
\sim \bar{h}$ does not hold in general. The further we increase $N$,
the more terms the specific subspace will contain ({\em e.g.} for
$N=3$, $s=-3$ we find 6 operators). This fact can be understood from
the explicit representation of the invariants (\ref{cons}). They have
partial degree $d_p(I_{2n+1}) = n+1$, or specifically for our example
from above $d_p(I_3)=2$. This means we can write $I_3/\s_3^2 =
\bar{I}_3$, where $\bar{I}_n$ denotes the same invariants (\ref{cons})
but in the variables $\bar{x}$.  Using these expressions the states
(\ref{sol2}) can be related to conformal states as
$$ Q^{(1)} \sim \bar{L}_{-1}^2 \sa ,\sc Q^{(2)} \sim \bar{L}_{-3}
L_{-1} \sb .$$

Similar for all spins $s$ in the subspace
$N=2$ we expect states of the kind $\bar{L}_{-3} L_{-n_1}
\dots L_{-n_k} \vert \phi_{1,2} \rangle$.  This is indeed the case,
since the generating function for the $N=2$ solutions is given by
$$ F_2 = \sum_{m,odd} \frac{q^{\frac 12 m (m-1)-2 m}}{(q)_m} = (1 +\q
+\q^2+\q^3) \c_{1,2}(q) \sb ,
$$ that is, it resembles exactly the structure of
$\c_{1,2}(\q)\c_{1,2}(q)$ up to second order in $\q$, but
additionally contains one term $\q^3 \c_{1,2}(q)$.

The presence of such higher level operators makes a direct link
between the parameter $N$ and the dimensions of the conformal fields
difficult.  Nevertheless increasing $N$ we can gradually build up the
whole space of operators in the massive theory.  The number of
operators at level $N$ is generated by $$ F_N\equiv
\sum_{m=1,odd}^{\infty} \frac{q^{\frac 12 m (m-1)-N m}}{(q)_m} \sb ,
$$ in the odd sector, and by
$$ G_N\equiv \sum_{m=0,even}^{\infty} \frac{q^{\frac 12 m (m-1)-N
    m}}{(q)_m} \sb ,$$ in the even sector.  These functions satisfy
the recursion relations \be F_N = F_{N-1} - \q^N G_{N-1}\sa , \sb G_N
= G_{N-1} + \q^N F_{N-1}\sb , \ee with the initial conditions $G_0 =
F_0 = \c_{1,2}(q)$. Because of the symmetry of the recursion relations
and the initial conditions, it follows that $F_N = G_N$.  This implies
that $F_N = (1+q^N) F_{N-1}$ which can be solved to give $F_N = G_N =
\prod_{k=1}^N (1+\q^k) F_0$, which resembles the character expansion
of $\c_{1,2}(\q) F_0$ up to order $N$.  Note that the whole space of
states is generated by $\lim_{N\to\infty} F_N$. This can be written as
\be F_\infty = \c_{1,2} (q) \c_{1,2}(\q) \sb ,\label{chir}\ee and
similarly for $G_\infty$.

A few comments follow, in order to clarify the meaning of these
expressions.  We find that $F_\infty$ and $G_\infty$ generate the
space of operators in this sector. They have the identical form of the
conformal expression of the space of descendent operators of the
primary field $\phi_{1,2}$.  Further note that in (\ref{chir}) the
characters of the two chiralities are separated. This is a main
feature of conformal field theories, where the right and left movers
can be treated independently.  Here in the massive theory the two
chiralities are present, but $q$ and $\q$ are {\em not independent
  variables}, and therefore such a factorisation can only be regarded
as a formal expression.

Finally we want to stress the fact that one obtains for finite $N$
some `finitized' expressions for the Virasoro characters. In this
approach they appear naturally in the counting procedure of states.
Note that the same expressions turn up in the study of corner transfer
matrices where the finitization is related to the fact that one has a
discrete system of finite size (see {\em e.g.} \cite{bc}). It would be
interesting to understand whether there is a deeper reason for this
relation.

The sector $w=0$ has been discussed in \cite{a14s}.  In this sector
the kinematical recursion relation maps the form factors onto zero,
$\F_n(\beta+i\pi,\beta,\beta_1,\dots,\beta_n) =0$.  It follows that
form factors with different particle number $n$ are {\em not} linked,
and therefore {\em any} kernel solution represents an acceptable form
factor from the point of view of the form factor equations.  The
analysis can be carried out in a similar way as for the sector
$w=\frac 12$. The result is that the even form factors generate the
descendent spaces of the energy operator, which scales in the
ultraviolet limit as $(\frac 12,\frac 12)$ and of the identity
operator $(0,0)$. The odd form factors generate the spaces of the
fermions $(\frac 12,0)$ and $(0,\frac 12)$. The ultraviolet limits of
the correlation functions can be calculated explicitly for these
operators and the scaling dimensions coincide with the corresponding
conformal operators.

The thermal Ising model has been examined also by various other means.
But, since it can be described as a free field theory, many methods
and results remain confined to this special model.  The interesting
fact about the method we have presented in this chapter is that it can
be straightforwardly applied also to interacting theories.  In fact
any massive integrable field theory having a description in terms of a
scattering matrix can be investigated in this way. We will give
several examples in the following sections.

\resection{Space of Operators in the massive Yang-Lee model\label{sec-yl}}

The scaling Yang-Lee model can be described as a massive integrable
field theory \cite{cm-yl}. Its on-mass-shell spectrum consists of a
single massive scalar particle of mass $m$.  Also the conformal field
theory $\M_{2,5}$ is quite simple containing only one primary field $
h=-\frac 15$ besides the identity operator.

The scattering matrix of the massive particles is determined by the
following two-particle amplitude\cite{cm-yl} \be S(\beta) \equiv
\frac{\tanh \frac 12(\beta+i \alpha )}{\tanh \frac 12 (\beta-i\alpha)}
\sb ,\label{smm}\ee with $\alpha= \frac 23 \pi$ \footnote{We keep
  whenever possible the notation general since we can use then some of
  the results we will obtain here for the analysis of the form factors
  of the models $\M_{2,2n+3}$, which will be discussed in section
  \protect\ref{sec-2n}. This is due to the fact that the $S$-matrix of
  the fundamental particle of these theories is given by
  (\protect\ref{smm}) with $\alpha= \frac 2{2n+1} \pi$.}.

The form factors have been analysed in \cite{Christe,smirnov13,YLZam}.
By defining the minimal two particle form factor as \be \F^{\rm
  min}_{11}(\beta) = (-i) \sinh \frac\beta 2 \, \exp \left \{
\int_0^\infty \frac{dx}x \frac{\cosh x (\frac 12 - \frac\alpha\pi)}
{\cosh \frac x2} \frac{\sin ^2 \frac{ x(i \pi-\beta)}{2\pi}}{\sinh x}
\right \} \sb ,\label{f32}\ee the parametrisation (\ref{paraf}) of the
form factors reads for this model as \be \F_n (\beta_1,\dots,\beta_n) =
H_n Q_n(\beta_1,\dots,\beta_n) \prod_{i<j} \frac{\F^{\rm
    min}(\beta_{i}-\beta_j)}{ (x_i+x_j) \sinh \frac 12 (
  \beta_{i}-\beta_j - i \alpha ) \sinh \frac 12 ( \beta_{i}-\beta_j +
  i \alpha )} \sb ,\label{parafyl}\ee where $Q_n$ is of the form
(\ref{ansatz}).  The constants $H_n$ in (\ref{parafyl}) are given by
\be H_n = C_0 \left ( 4 \cos ^2 \frac \alpha 2\, \sin \alpha \right
)^{ \frac n 2} (\zeta(i\pi)\pi)^{\frac{(n-1)^2-1}{2}}\, i^{n^2} \sb
.\label{hn}\ee

The form factor equations have been reduced through this
parametrisation to two coupled recursive relations for $Q_n$
\cite{YLZam}, namely \be Q_{n+2} (-x,x,x_1,\dots,x_n) =
D_n(x,x_1,\dots,x_n) Q_n (x_1,\dots ,x_n) \label{recursive} \sb ,\ee
with the recursion function  \be D_n(x,x_1,\dots,x_n) =
\sum_{k=1}^n \sum_{m=1,odd}^k [m]\, x^{2(n-k)+m}
\sigma_{k}^{(n)}\sigma_{k-m}^{(n)} (-1)^{k+1} \sb ,
\label{D_n}
\ee wherein the symbol $[l]$ is defined by \be [l]\equiv\frac{\sin (
  l\alpha)}{\sin\alpha} \sb .\label{f37}\ee The second recursion
relation derives from the bound state axiom (\ref{bounds}) and reads
as \be Q_{n+1}(x\omega^{\frac 12}, x\omega^{-\frac
  12},x_1,\dots,x_{n-1}) = x \prod_{i=1}^{n-1} (x+x_i)
\,Q_n(x,x_1,\dots,x_{n-1}) \sb .\label{bsyl}\ee

The recursion relation (\ref{recursive}) has been studied in
\cite{a10} for the primary massive operators.  It admits an infinite
number of solutions. A basis in the solution space is given by the
polynomials \be Q_n= \Q_{n}(k) \equiv \vert\vert M_{ij}(k) \vert\vert
\sa ,\sa k \in \Z \sb ,\label{sol} \ee where $M_{ij}(k)$ is an
$(n-1)\times (n-1)$ matrix whose entries are \be M_{ij}(k) =
\s_{2i-j}\, [i-j+k] \sb .
\label{element}
\ee Enforcing in addition to (\ref{recursive}) also the bound state
recursion relation (\ref{bsyl}) only {\em one} consistent solution
exists, which is given by $\Q_n(1)$ \cite{a11}. In the ultraviolet
limit the operator corresponding to this solution scales as the
conformal primary operator $\phi_{1,2}$ of the minimal model $\M_{2,5}$
\cite{YLZam}.  For $w=\frac 12$ the equation (\ref{recursive}) admits
only one single solution \cite{lash}, which is not consistent with the
bound state equation (\ref{bsyl}).

We are interested in the analysis of the space of descendent operators
of this model. From (\ref{ansatz}) we find that the functions $Q_n$
take the form $Q_n(x_1,\dots,x_n)=\s_n^{-N} P_n(x_1,\dots,x_n)$.  We
start by analysing the space of chiral descendents, $N=0$.  Note, that
in this model even and odd form factors are coupled due to the bound
state equation (\ref{bsyl}).  The method is in principle the same as
in the Ising model.  We solve the relations (\ref{recursive}) and
(\ref{bsyl}) recursively using the property that the space of
solutions at level $n$ is given by the number of solutions at level
$n-1$ plus the dimension of the kernel of the combined recursion
relations. In table \ref{tab-1} we have written down the degree for
the polynomials $P_n$ required from Lorentz invariance and the
dimension of the kernel of the recursion relations.

Let us now count
the independent solutions for various spin levels.  For $s=1$ the only
solution is that generated by $P_1= \s_1$, which corresponds exactly
to the level 1 descendent of the primary field $\phi$ of the Yang-Lee
model. For spin 2 we have two possible solutions, generated by the
polynomials
$$ P_1= \s_1^2\sa , \sb P_2 =\K_2 \sb .$$ Similarly higher spin values
can be investigated. From table \ref{tab-1} we find that a kernel
solution for $n=3$ occurs only for spin $s \geq 6$. Let us examine the
solution space for the particular spin value $s=6$. The solutions are
\bea P_1 &=& \s_1^6 \sb ,\nonumber \\ P_2 &=& \s_1^4\K_2,\, \s_2
\s_1^2\K_2,\, \s_2^2 \K_2\sb , \nonumber\\ P_3 &=& \K_3 \sb ,\eea that
is we have 5 independent operators.

We compare these values with the dimensions of the spaces of
descendent operators in conformal field theory.  The number of
descendent operators is given by the character expansions,
\begin{table}[tbh]
\begin{center}\begin{tabular}{||c|c|c||} \hline $ n$&
  $deg (Q_n^{(s)})=\frac 12 n(n-1)+s$ & $deg \K = \frac 32 n(n-1)
  n\geq 2 $
\\
\hline
$1$ &$ 0+s$ &$ 0$ \\
$2$ &$1+ s$&$3  $\\
$3$ &$3 +s$&$ 9 $\\
$4$ &$6 +s$&$ 18$ \\
$5$ &$10+s$ &$30$  \\
$6$ &$15+s$ &$45$  \\
$7$ &$21+s$ &$63$  \\
\hline\end{tabular}\end{center}
\caption{Comparing the total degrees of the polynomials $P_n$ for the
  case $N=0$ with the kernel of the recursion relations for spin $s$
  and level $n$.}
\label{tab-1}
\end{table}
\bea \chi_{1,1} &=& 1+q^2+q^3+q^4+2 q^5 +2 q^6 +O(q^7) \sb
,\nonumber\\ \chi_{1,3} &=& 1+q+q^2+q^3+2 q^4+2 q^5 +3 q^6 +O(q^7) \sb
.\eea Summing up these values we find that the Yang-Lee CFT contains 1
descendent operator at spin 1, 2 operators at spin 2 and 5 operators
at spin 6. We compare these values with the ones obtained in the
perturbed model and find that they coincide.

In table \ref{tab-2} we have collected the number of independent
solutions of the recursion relations using the notation of partitions
introduced in (\ref{parts}). The counting can now be generalised to
arbitrary values of $s$. A new kernel solution will contribute,
whenever $s=n(n-1)$.  Therefore the number of solutions is generated
by \be F_1 = \sum_{n=0}^\infty \frac {q^{n (n-1)}}{(q)_n}\sb , \ee
where we introduce the general definition, \be F_N \equiv
\sum_{n=0}^\infty \frac {q^{n^2 - N n }}{(q)_n}\sb . \label{all0} \ee

\begin{table}[tbh]
\begin{center}\begin{tabular}{||c|l||} \hline
 $ s=0$ & $ \P(1,0)$ \\
 $ s=1$ & $ \P(1,1)$ \\
 $ s=2$ & $ \P(1,2) + \P(2,0)$ \\
 $ s=3$ & $ \P(1,3) + \P(2,1)$ \\
        &     $ \vdots$   \\
 $ s=6$ & $ \P(1,6) + \P(2,4) + \P(3,0)$ \\
\hline\end{tabular}\end{center}
\caption{Number of independent solutions of the recursion equations
  for several spin values $s$.}
\label{tab-2}
\end{table}

For the Yang-Lee model the Rogers-Ramanujan identities (see {\em e.g.}
\cite{andrews,rademacher}) relate the so-called \cite{kedem}
`fermionic' sum expressions to the `bosonic' form of
\cite{rocha,Christe} as \bea \c_{1,1}^{(2,5)}(q) = \prod_{i=0}^\infty
\frac 1{(1-q^{2+5 i}) (1-q^{3+5 i})} &=& F_{-1} \sb ,\nonumber\\
\c_{1,2}^{(2,5)}(q) = \prod_{i=0}^\infty \frac 1{(1-q^{1+5 i})
  (1-q^{4+5 i})} &=& F_0 \sb .\eea By a re-summation one can show that
$$ F_1 = F_0 + F_{-1} =\c_{1,1}(q)+\c_{1,2}(q) \sb ,$$ which means
that the number of chiral descendent operators at each level is the
same in the conformal and the perturbed Yang-Lee model. It is
interesting to note that we have found the {\em sum} of the two
characters, and not the two Virasoro irreducible representations
modules separately. This is expected from the physical point of view,
since in the perturbed model a mixing of the two spaces occurs.  This
has been analysed by Zamolodchikov \cite{Zam}, by determining the
conservation laws of the perturbed theory. The simplest such
conservation law, is given by
$$\partial_{ \bar{z}} T \sim \partial_z \phi_{1,2} \sb ,$$ where $T$
denotes the holomorphic component of the energy momentum tensor, which
is an element of the Virasoro irreducible representation of the
identity operator in the conformal model. This relation expresses the
continuity equation for the stress energy tensor in the massive model.

Next we investigate the solutions corresponding to $Q_n = \frac
1{\s_n} P_n$, {\em i.e.} $N=1$ in (\ref{ansatz}).  Let us repeat the
counting as we have done for the primary operators. The total degree
of the polynomials $P_n$ is now given by $deg(P_n)= \frac 12 n(n-1) +n
+s$.  The first few solutions are \be \ba{ll} s=-1 & P_1 =const.\sb ,
\\ s=0 & P_1 = \s_1 , \,P_2=\K_2 \sb ,\\ s=1 & P_1 = \s_1^2 ,\,
P_2=\s_1 \K_2\sb , \\ s=2 & P_1 = \s_1^3 ,\, P_2=\s_1^2 \K_2,
\,P_2=\s_2 \K_2 \sb .\ea \ee As in the Ising model this amounts in an
over-counting of the states, since the chiral operators appear in the
subspace $N=1$ as solutions proportional to $\s_n$, {\em i.e.} for
which $P_n \sim \s_n$.

The generating function of the $N=1$ states is given by $F_2$.  In
order to find the structure of this subspace, we subtract the chiral
operators using the $q$-sum relation
$$F_2-F_1 = \q F_0 = \q \c_{13} (q) \sb .$$ The structure of the $N=1$
space coincides with that of the descendent operators of the conformal
scalar primary fields $\phi_{1,1}$ and $\phi_{1,3}$ for $\bar{h}=1$.
In CFT there are only
states proportional to $\c_{1,3}(q)$ since
$\bar{L}_{-1} \vert I \rangle =0$. We find therefore that the subspace
$N=1$ in the massive model corresponds to the space
$\bar{h}=1+\bar{h}_0$ of the conformal model, where $h_0$ stands for
the right chiral dimension of the conformal primary fields.

Using the generating function (\ref{all0}) one can show that the
number of operators in the subspace $N$, is generated by $F_{N+1}$.
This function satisfies the recursion identity \be F_N = F_{N-1} +
\q^{N-1} F_{N-2} \sb ,
\label{recyl} \ee with the initial condition of $F_0 = \c_{1,3}(q)$
and $F_{-1} = \c_{1,1}(q)$. It has been shown by I. Schur (see {\em
  e.g.} \cite{rademacher}) that the solution to this problem is given
by \be F_N = D_1^{(N-1)}(\q) \c_{1,3} (q) + D_2^{(N-2)}(\q)
\c_{1,1}\sb , \ee with \be D_1^{(N)}(\q) = \sum_{\mu=-\infty}^{\infty}
(-1)^\mu \q^{\frac \mu 2 (5 \mu +1)} \left [ \ba{c} N+1 \\ \, [ \frac
{N+1-5 \mu}{2}] \ea \right ] \sb ,\ee \be D_2^{(N)}(\q) =
\sum_{\mu=-\infty}^{\infty} (-1)^\mu \q^{\frac \mu 2 (5 \mu -3)} \left
[ \ba{c} N+1 \\ 1+ [ \frac {N+1-5 \mu}{2} ] \ea \right ] \sb ,\ee
which are finitized forms of the characters $\c_{1,3}(\q)$ and
$\c_{1,1}(\q)$ respectively. From the recursion relation (\ref{recyl})
it follows that $F_N = F_{N-1} (mod\, \q^N)$, and therefore we find
that $F_N$ resembles the partition function of the Yang-Lee model up
to order $\q^N$.  Similarly as in the Ising model we can write the
formal identity \be \lim_{N\to \infty} F(N) = Z_{2,5} = \c_{1,3}(q)
\c_{1,3}(\q) + \c_{1,1} (q) \c_{1,1} (\q) \sb .\ee

There are several new features in this model with respect to the Ising
model. First in the form factor approach we cannot disentangle members
of the Virasoro irreducible representations of the identity operator
and the primary field. As we have explained this is due to the fact,
that a mixing of these Virasoro irreducible representations occurs in
the massive theory. On the other hand if the structure of the spaces
are isomorphic one should be able to find a unique mapping of the
operators.

Actually a similar structure appears in the Ising model in the sector
$w=0$, which contains the descendent operators of the massive scalar
primary fields $\phi_{1,3}$ and $\phi_{1,1}$.  Also there the same
argument of the mixing of operators holds.The special fact in the
Ising model is that for the sector $w=0$ the UV dimensions of the
operators can be calculated explicitly. Therefore it is possible to
assign in a unique manner the form factors to specific conformal
operators. It has been shown in \cite{a14s} that in the Ising model
($w=0$) chiral operators in the descendent space of the identity
operator have form factors
which are proportional to $\s_n$ while the others form the chiral
descendents of the primary field $\phi_{1,3}$.

It is tempting to conjecture a similar structure in the Yang-Lee
model. This is due to the fact that the space of solutions with
$P_n \not\sim \s_n$ are given by
$$\sum_n \frac{q^{n(n-1)}}{(q)_{n-1}} = \c_{1,1}(q)\sb .$$ This seems
to be the opposite situation as in the Ising model, but one has to
recall that in the Yang-Lee model the field $\phi_{1,3}$ has {\em
  negative} scaling dimension, and therefore behaves less singular as
the identity operator in the UV limit. Unfortunately we are not able
to calculate the UV dimensions exactly in this model, and therefore
this conjecture cannot be confirmed.

Finally we want to discuss the relation of the space of operators to
the conserved charges of the model.  In \cite{Christe} P.  Christe has
constructed descendent operators of the primary field. His approach
was to modify the construction of \cite{CM} for the Ising model to the
case of the Yang-Lee model, {\em i.e.} he constructed polynomial
invariants which satisfied both recursion relations (\ref{recursive})
and (\ref{bsyl}).  He found invariants of spin values $s=1,5\,
mod\,(6)$.  They correspond to the infinite set of integrals of
motions of the massive theory. Note that in the Ising model
polynomials formed the corresponding invariants can be mapped
one-to-one to the fields of the conformal Virasoro irreducible
representation. Obviously this is not the case for the Yang-Lee model,
since the operators of the Virasoro irreducible representation of
$\phi_{1,2}$ are generated by $L_{5 n+1},L_{5 n+4}$.  Constructing
derivative operators through polynomial invariants one only finds a
{\em sub-space} of the descendents of the conformal operator
$\phi_{1,2}$. This implies that the conserved charges do {\em not}
generate the whole space of operators.  This feature generalises to
all further models we will consider. The Ising model is an exception
since it can be considered as a free field theory.

\resection{The model $\M_{3,5}$\label{sec-35}}

As a further simple model whose on-shell spectrum contains only one
particle is the $\phi_{1,3}$ perturbation of $\M_{3,5}$.  Let us
recall the basic properties of the conformal model.  It is a
non-unitary minimal model, with the primary operators given through
the Kac table \be
\begin{array}{|c|c|}\hline
 \frac{3}{4} & 0\\ \hline
 \frac{1}{5} & -\frac{1}{20} \\ \hline
 -\frac{1}{20} & \frac{1}{5} \\ \hline
 0 & \frac{3}{4} \\
\hline
\end{array} \sb .
\label{tab35}
\ee
The characters of these operators
are \cite{rocha,kedem},
\bea \c_{1,1}(q) &=&\sum_{n=0,\, even} \frac{ q^{ \frac{n(n+2)}{4}}
  }{(q)_n}= \sum_{\mu=-\infty}^{\infty} q^{\mu (15 \mu + 2)}-
q^{(3\mu+1)(5\mu+1)}\sb , \nonumber \\ \c_{1,2}(q) & =& \sum_{n=0,\,
  even} \frac{ q^{ \frac{n^2}{4}} }{(q)_n}=
\sum_{\mu=-\infty}^{\infty} q^{\mu (15 \mu -1)}- q^{(3\mu+1)(5\mu+2)}
\sb ,\nonumber \\ \c_{1,3}(q) & =& \sum_{n=0,\, odd} \frac{ q^{
    \frac{n^2-1}{4}} }{(q)_n}= \sum_{\mu=-\infty}^{\infty} q^{\mu (15
  \mu -4)}- q^{(3\mu+1)(5\mu+3)} \sb ,\nonumber \\ \c_{1,4}(q) &=&
\sum_{n=0,\, odd} \frac{ q^{ \frac{n(n+2)-3}{4}} }{(q)_n}=
\sum_{\mu=-\infty}^{\infty} q^{\mu (15 \mu - 7)}- q^{(3\mu+1)(5\mu+4)}
\sb .\label{tab-35} \eea We have collected the bosonic and the
fermionic expressions for later convenience.

Perturbing the conformal model in the direction of the $\phi_{1,3}$
operator one obtains a massive field theory, containing one
asymptotic particle without a bound state. The interaction of the
asymptotic states is given through the $S$-matrix \cite{smi13}
$$S(\beta)\,=\,-i\,\tanh\frac{1}{2}\left(\beta -
i\frac{\pi}{2}\right)\sb .$$ An interesting novel fact with respect to
the previous models is, that these particles have non trivial
statistics, since $\lim_{\beta\rightarrow \pm\infty} S(\beta)\,=\,\mp
i$, and are in fact related to the operator $\phi_{2,1} = \phi_{1,4}$
of the model \cite{smiCMP}. Because they are spin $\frac 14$ particles
they also have $w_{ij} = \mp \frac 14$ depending which chirality of
the operator we choose (or equivalently if we define $w_{ij}$ through
$\pm \infty$ in rapidity space).

 The minimal two particle form factor can be determined as \cite{DM}
 \be \F^{\rm min}(\beta)= \sinh\frac{\beta}{2}\, \prod_{k=0}^{\infty}
 \frac{\Gamma\left(k+\frac 14-i\frac{\beta}{2\pi}\right)
   \Gamma\left(k+\frac 54+i\frac{\beta}{2\pi}\right)}
 {\Gamma\left(k+\frac34-i\frac{\beta}{2\pi}\right) \Gamma\left(k+\frac
   74+i\frac{\beta}{2\pi}\right)} \sb .\ee Since there are no bound
 states the parameterisation of the form factor is given by \be
 \F_{n}(\beta_1,\ldots,\beta_{n})= H_{n}\, Q_{n}(x_1,\ldots,x_{n})\,
 \s_n^{\frac n4} \s_n^{-\frac 12 \delta_{w,0}} \prod_{i<j}
 \frac{\F^{\rm min}(\beta_{ij})} {x_i+x_j}\sb , \label{arm35}\ee where
 we define the normalisation constants $H_n$ such that they satisfy
 \be H_{n+2} = \left ( \frac{2i}{\pi}\right )^n \frac 1{\F^{\rm min}(i
   \pi)} H_n \sb .\ee The fractional power of $\s_n$ in (\ref{arm35})
 in the sector $w=0$ is needed in order to satisfy (\ref{wat2}). As a
 consequence we will find that operators corresponding to {\em odd}
 form factors will have non-zero spin, namely the same as the
 asymptotic particles, $s=\pm \frac 14$.

 Using the parametrisation (\ref{arm35}) and taking into account the
 statistics of the asymptotic states, the recursion relation
 (\ref{kin}) reduces to \be
 Q_{n+2}(-x,x,x_1,\dots,x_{2n})\,=\,(-x)^{\delta_{w,0}}\, \,
 D_{n}(x,x_1,\dots,x_{n})\, Q_{n}(x_1,\ldots,x_{2n})\sb ,
\label{kin35}
\ee where \be D_{n}(x,x_1,\dots,x_{n})\,=\,\sum_{k=0}^{n} (e^{(n-k)
  \frac{i \pi}{2}} - e^{2 i \pi w} e^{- (n-k)\frac{i \pi}{2}}) x^{n-k}
\sigma_k^n \sb .\label{recf35}\ee

We first analyse the sector $w=0$.  Even and odd form factors
are not linked by the recursive equations, and we start by examining
the even ones.  As before we use the parametrisation (\ref{ansatz}) to
introduce a grading in the space of operators. We have $\gamma=0$
since the fractional powers of $\s_n$ have already been taken in
account in (\ref{arm35}).

Since we have carried out the counting technique already in detail for
the Ising and the Yang-Lee model we will merely state the results.
For chiral operators ($N=0$) the degree of $P_n$ is given by $deg(P_n)
= \frac{n^2}{4}+s $.  For spin $s=0$ there exists only one solution
which was found in \cite{DM}, and corresponds to the perturbation of
the scalar primary field $\phi_{1,3}$. Its form factors are given by
$Q_n = ||\s_{4 i-2 j-1}||$, $n=2,4,\dots$ in (\ref{arm35}), where the
dimension of this determinant is $\frac n2$.

For spin $s=0,1,2$ one finds $1,1,3$ solutions respectively, while for
general spin values the generating function for the number of
independent solutions is \be F_1^e = \sum_{n,\, even} \frac{
  q^{\frac{n(n-2)}{4}}}{(q)_n}\sb .  \ee By shifting the summation
index and using the fermionic sum expressions of the characters
(\ref{tab-35}) it is not difficult to show that \be F_1^e =
\c_{1,1}(q) + \c_{1,3}(q) \sb .  \ee

We want to analyse the full space of operators, and therefore increase
the value $N$ in (\ref{ansatz}) by integer powers.  The generating
function for states at level $N$ is given by \be F_N^e \equiv
\sum_{n,\, even} \frac{ q^{\frac{n(n+2)}{4}-Nn}}{(q)_n}\sb .
\label{ggg35}\ee Let us further define \be G_N^o \equiv \sum_{n,\,
  odd} \frac{ q^{\frac{n^2-1}{4}-N n}}{(q)_n} \sb ,\label{ggf35}\ee
then these functions satisfy the recursion identities \be \ba{l}F_N^e
=F_{N-1}^e + \q^{N-1} G_{N-1}^o \sb ,\\ G_N ^o= G_{N-1}^o + \q^{N}
F_N^e \sb .
\label{gf35}\ea \ee In order to separate formally
the chiral sectors, we express these infinite sums in terms of the
characters. From the recursion relations it is clear that $F_N^e$ and
$G_N^o$ can be written as polynomials in $\q$ multiplied by the
characters, which are the initial conditions of the recursion
relations, namely $F_0^e = \c_{1,1}(q)$ and $G_0^o = \c_{1,3}(q)$.
Expanding for the first few values of $N$ one finds indeed that
$$F_N^e = \c_{1,1} (q) \c_{1,1}(\q) + \c_{1,3}(q) \c_{1,3}(\q) \sa
(mod\, \q^N )\sb . $$

The general solution of the recursion relations can be cast into the
form \bea F_N^e &=& D_1^{(N)}(\q) \c_{1,1}(q) + D_3^{(N)}(\q)
\c_{1,3}(q)\sb , \nonumber\\ G_N^o &=& D_4^{(N)}(\q) \c_{1,1}(q) +
D_2^{(N)}(\q) \c_{1,3}(q) \sb ,\eea with \bea D_1^{(N)}(\q) &=&
\sum_{\mu=-\infty}^{\infty} \q^{ \mu (15 \mu +2)} \left [ \ba{c} 2 N
\\ N-5 \mu \ea \right ]- \q^{(3 \mu+1) (5 \mu +1)} \left [ \ba{c} 2 N
\\ N-1-5 \mu \ea \right ] \to \c_{1,1}(\q) \nonumber \sb ,\\
D_2^{(N)}(\q) &=& \sum_{\mu=-\infty}^{\infty} \q^{ \mu (15 \mu +1)}
\left [ \ba{c} 2 N \\ N-5 \mu \ea \right ]- \q^{(3 \mu+1) (5 \mu +2)}
  \left [ \ba{c} 2 N \\ N-3-5 \mu \ea \right ] \to \c_{1,2}(\q) \sb
    ,\\ D_3^{(N)}(\q) &=& \sum_{\mu=-\infty}^{\infty} \q^{ \mu (15 \mu
      -4)} \left [ \ba{c} 2 N
  \\ N-1-5 \mu \ea \right ]- \q^{(3 \mu+1) (5 \mu +3)} \left [ \ba{c}
  2 N \\ N-2-5 \mu \ea \right ] \to \c_{1,3}(\q)\sb ,\nonumber \\
  D_4^{(N)}(\q) &=& \sum_{\mu=-\infty}^{\infty} \q^{ \mu (15 \mu +7
    \mu +1)} \left [ \ba{c} 2 N \\ N-1-5 \mu \ea \right ]- \q^{(3
    \mu+2) (5 \mu +1)+1} \left [ \ba{c} 2 N \\ N-2-5 \mu \ea \right ]
  \to \q \c_{1,4}(\q) \sb , \nonumber\eea where the limits are
  understood as $N \to \infty$.

  While $F_N^e$ generates the even part of the partition function as
  we expect, it seems that the function $G_N^o$ lacks a physical
  interpretation in this context. Its importance will be clarified
  in the following.

Let us introduce a compact notation in order to identify the operators
corresponding to a specific set of form factors. If the space of
operators in a certain sector can be decomposed as $\c_{r,s}(q)
\c_{r',s'}(\q)$ we will call the corresponding chiral ({\em i.e.} with
the lowest possible divergence of the form factors) operator with spin
$s=h_{r,s}-h_{r',s'}$, as the perturbed {\em primary} field
$(h_{r,s},h_{r',s'})$. In this notation the space of even form factors
in the sector $w=0$ contains the descendents of the primary fields
$(\frac 15, \frac 15)$ and $(0,0)$.

The analysis of the even form factors is complete, and we turn to the
odd sector. From (\ref{prop2}) and (\ref{arm35}) we find that it will
contain particles of spin $-\frac 14 +\Z$. By applying the counting
method we find that the space of chiral descendents is given by \be
F_1^o\sum_{n,odd} \frac{q^{\frac{n(n-2)}{4}}}{(q)_n} = q^{-\frac 14}
\c_{1,2}(q) +q^{\frac 34}\c_{1,4}(q) \ee which indicates that
$\phi_{1,4}$ does not mix with the primary operator $\phi_{1,2}$ but
with its descendent at level one.  Compare this situation with the
table of conformal dimensions of the model (\ref{tab35}).  The
dimension of the field $\phi_{1,4}$ ($h=\frac 34$) is by far `closer'
to that of $L_{-1}\phi_{1,2}$ ($h=\frac {19}{20}$) than to that of the
primary operator itself ($h=-\frac 1{20}$).  Therefore this result is
not surprising. On the other hand this indicates that when studying
more complex systems, where the dimensions of the primary operators
differ by several units (as it is the case for example for the unitary
minimal models), this counting procedure might become rather complex,
since different primary operators will mix with descendent operators
at various levels.

For general $N$ we use again generating functions $F_N$ and $G_N$ now
defined as \be F_N^o \equiv \sum_{n,\, odd} \frac{
  q^{\frac{n(n+2)}{4}-Nn}}{(q)_n}\sb {\rm and}\sb G_N^e \equiv
\sum_{n,\, even} \frac{ q^{\frac{n^2-1}{4}-N n}}{(q)_n} \sb .\ee
Observe that they have the same functional form as the generating
functions for the even sector (\ref{ggg35}) and (\ref{ggf35}), but the
summation over odd and even indices is interchanged. They satisfy the
{\em same} recursion relations (\ref{gf35}) but with the initial
conditions $F_0^o = q^{\frac 34} \c_{1,4}(q)$ and $G_0^e=q^{-\frac
  14}\c_{1,2} $. This implies that the space of operators can be
decomposed as \be F_\infty^o = q^{-\frac 14} \c_{1,2} (q) \c_{1,3}
(\q) + q^{\frac 34} \c_{1,4} (q) \c_{1,1} (\q)\sb .\ee

We have calculated the form factors of the primary operators
explicitly, and they are given by the polynomials $Q_n$ in
(\ref{arm35}) \be Q_n = ||\s_{4i-2j-1}||^{<\frac{n-1}2>}\sb ,\sa
n=1,3,\dots, \ee for the field $(-\frac1{20},\frac 15)$ where the
upper index of the determinant indicates the dimension of the
corresponding matrix.  For the operator $(\frac 34,0)$ the polynomials
are \be Q_1=\s_1\sb {\rm and}\sb Q_n = \s_n ||
\s_{4i-2j}||^{<\frac{n-3}2>} \sa,\sa n=3,5,\dots \sb .\ee

We now turn our attention to the sector $w=\frac 12$.  The recursion
function (\ref{recf35}) takes the form \be D_n = \sum \s_k x^{n-k}
\cos \frac \pi 2 (n-k)\sb . \ee The counting argument determines the
space of solutions, as being generated by \be G_0^e = \sum_{n, even}
\frac {q^{\frac{n^2}{4}}}{(q)_n} = \c_{1,2}(q) \sb .\ee In order to
understand the structure of the space of operators we also analyse
explicitly the terms with $N=1$.  It is given by \be G_1^e = \sum_{n,
  even} \frac {q^{\frac{n^2}{4}-1}}{(q)_n} = (1+\q) \c_{1,2}(q)
+\c_{1,4}(q)\sb , \ee that is $\phi_{1,4}$ does not mix with the
primary operator $\phi_{1,2}$ but again with its first descendent. We
have drawn a picture of the perturbed space of operators in figure
\ref{fig-35}.

\begin{figure}[htb]
\begin{picture}(400,250)(-200,0)
\thicklines
\put(-200,10){\line(1,0){400}}
\put(0,0){\line(0,1){220}}
\put(0,0){\line(1,1){220}}
\put(0,0){\line(-1,1){220}}
\thinlines
\put(0,80){\line(1,1){140}}
\put(0,80){\line(-1,1){140}}
\multiput(0,0)(40,40){5}{\circle*{4}}
\multiput(-40,40)(40,40){4}{\circle*{4}}
\multiput(-80,80)(40,40){3}{\circle*{4}}
\multiput(-120,120)(40,40){2}{\circle*{4}}
\multiput(-160,160)(40,40){1}{\circle*{4}}
\put(-220,200){$N$}
\put(-4,200){\makebox(0,0)[r]{$\Delta = h + \bar{h}$}}
\multiput(-160,5)(40,0){9}{\line(0,1){10}}
\put(-160,-7){\makebox(0,0){$-4$}}
\put(-120,-7){\makebox(0,0){$-3$}}
\put(-80,-7){\makebox(0,0){$-2$}}
\put(-40,-7){\makebox(0,0){$-1$}}
\put(40,-7){\makebox(0,0){$1$}}
\put(80,-7){\makebox(0,0){$2$}}
\put(120,-7){\makebox(0,0){$3$}}
\put(160,-7){\makebox(0,0){$4$}}
\put(195,13){$s$}
\put(6,0){\footnotesize $\phi_{1,2}$}
\put(46,35){\footnotesize $L_{-1} \phi_{1,2}$}
\put(86,75){\footnotesize $L_{-1}^2 \phi_{1,2}$}
\put(126,115){\footnotesize $L_{-1}^3 \phi_{1,2}$, $L_{-3} \phi_{1,2}$}
\put(166,155){\makebox(0,0)[l]{\footnotesize\shortstack{
 $L_{-1}^4 \phi_{1,2}$\\ $L_{-3} L_{-1}\phi_{1,2}$\\ $L_{-4} \phi_{1,2}$}}}
\put(-34,35){\footnotesize $\bar{L}_{-1} \phi_{1,2}$}
\put(6,75){\makebox(0,0)[l]{\footnotesize\shortstack[l]{
 $\bar{L}_{-1} L_{-1} \phi_{1,2}$ \\ $\phi_{1,4}$}}}
\put(46,115){\makebox(0,0)[l]{\footnotesize\shortstack[l]{
 $\bar{L}_{-1} L_{-1}^2 \phi_{1,2}$\\ $L_{-1} \phi_{1,4}$}}}
\put(-74,75){\footnotesize $\bar{L}_{-1}^2 \phi_{1,2}$}
\put(-34,115){\footnotesize\shortstack[l]{
$\bar{L}_{-1}^2 L_{-1}\phi_{1,2}$\\ $\bar{L}_{-1} \phi_{1,4}$}}
\put(-114,115){\makebox(0,0)[l]{\footnotesize\shortstack{
 $\bar{L}_{-3} \phi_{1,2}$\\ $\sa\,\bar{L}_{-1}^3 \phi_{1,2}$}}}
\end{picture}
\caption{Space of descendent operators in the massive model
  $\M_{3,5} +\phi_{1,3}$ for the sector $w=\frac 12$.}
\label{fig-35}
\end{figure}

The full space of operators is generated by $G_n^e$ which becomes in the
limit $n\to\infty$ \be G_\infty^e = \c_{1,2}(q)
\c_{1,2}(\q)+q\q\,\c_{1,4}(q) \c_{1,4}(\q) \sb .\label{above} \ee The
form factors of the primary operator $(-\frac 1{20},-\frac1{20})$ can
be calculated explicitly, and are given by
$$Q_n=|| \s_{4i-2j} ||^{<\frac n2 -1>}\sa ,\sa n=2,4,\dots,$$ in
(\ref{arm35}).  In this example it becomes clear that the expressions
describing the content of operators in a specific sector in terms of
the characters have to be regarded as formal ones. Especially in
(\ref{above}) there appears the term $q\q$ which multiplies the
characters of the field $\phi_{1,4}$. If we cancel them, we would not
alter (\ref{above}), but the expression would no longer reflect the
actual structure of the space of operators in this sector.

Finally let us summerise the analysis of the odd form factors in the
sector $w=\frac 12$. We find that the space is structured as \be
G^o_{\infty} =q^{\frac 14} \c_{1,3} (q) \c_{1,2} (\q) + q^{-\frac 34}
\c_{1,1} (q)\c_{1,4} (\q) \sb .\ee The form factors of the primary
fields are
$$Q_n = ||\s_{4i-2j-1}||^{<\frac{n-1}2>}\sa ,\sa n=1,3,\dots, $$ for
the operator $(\frac 15,-\frac 1{20})$ and
$$Q_1=\bar{\s}_1 \sb{\rm and}\sb Q_n= ||\s_{4i-2j+1}||^{<\frac{n-3}2>}
\sa,\sa n=3,5,\dots ,$$ for the operator $(0,\frac 34)$.

Let us discuss the physical implications of the above analysis. The
massive model $\M_{3,5}+\phi_{1,3}$ contains the following primary
operators: the scalar fields $(\frac 15,\frac 15)$ and $(0,0)$ and the
para-fermionic fields ( $s=-\frac 14 +\Z$), namely $(-\frac
1{20},\frac 15)$, $(\frac 34,0)$ in the sector $w=0$. In the sector
$w=\frac 12$ we find the fields $(-\frac 1{20}, -\frac 1{20})$,
$(\frac 34,\frac 34)$ and the spin $s=\frac 14+\Z$ operators $(\frac
15,-\frac 1{20})$, $(0,\frac 34)$. We want to stress the fact, that we
are {\em not} able to establish a direct link between the form factors
of these massive operators and their conformal counterparts, but have
rather defined those operators through the degeneracies of their
descendent spaces.

An interesting feature of this model is the appearance of the
para-fermionic operators. Their spin is determined by the spin of the
asymptotic states. This causes that only specific combinations of
chiral dimensions can appear in the massive model. Thus in contrast to
the space of operators in the conformal theory where arbitrary
combinations of right and left dimensions can occur for a theory
defined on the real line.

We have introduced the chiralities in the massive model through the
variables $x$ and $\bar{x}=\frac 1x$. Therefore, in order to have a
consistent description we should be able to transform the form factors
of the operator $(-\frac 1{20},\frac 15)$ to those of the operator
$(\frac 15,-\frac 1{20})$ by the transformation $x\to \bar{x}$ (and
similarly for the other pair of para-fermionic fields).  Let us
analyse the form factors of $(-\frac 1{20},\frac 15)$, given by \be
\F_n^{(-\frac 1{20},\frac 15)} = H_n \s_n^{\frac n2 -\frac 12}
||\s_{4i-2j}||^{<\frac{n-1}2>} \, \prod\frac{F^{\rm min}
  (\beta_{ij})}{(x_i+x_j)} \sb .\ee We define
$\bar{\F}_n(\beta_1,\dots,\beta_n) =\F_n(-\beta_1,\dots,-\beta_n)$.
Using the relation (\ref{sbar}) we find for the determinant \be ||
\bar{\s}_{4i-2j}||^{<\frac{n-1}2>} = \s_n^{-\frac{n-1}2}
||\s_{4i-2j-1}||^{<\frac{n-1}2>} \sb ,\ee and therefore \be
\bar{\F}_n^{(-\frac 1{20},\frac 15)}= \F_n^{(\frac 15,-\frac 1{20})}
\sb .\ee Analogously one can show that $\bar{\F}_n^{(0,\frac 34)}=
\F_n^{(\frac 34,0)}$.

Related to this fact is the following observation.  In calculating the
form factors, we had to choose which chirality {\em i.e.} which limit
of the $S$-matrix to select in the analytic continuation process. We
chose the limit $\beta \to \infty$. Suppose we would choose $S(-\infty) =
+i$. Repeating the same analysis as before, one finds that the scalar
fields remain in the respective sectors, while the spin-dependent
fields change place. This is as expected, since now the chirality of
the spin is now linked to the {\em right} component of the field
dimensions.

A final remark concerns the scalar operators of the theory. We have
found $\phi_{1,1}$ and $\phi_{1,3}$ in the sector $w=0$ while we found
$\phi_{1,2}$ and $\phi_{1,4}$ in the sector $w=\frac 12$. It is
interesting to note that these locality properties are in accordance
with the operator-product expansions of the correspondent conformal
 operators with the field $\phi_{1,4}$. It has been observed
previously \cite{smiCMP} that the field $\phi_{2,1}$ (which for
$\M_{3,5}$ is the same as $\phi_{1,4}$) is deeply connected to the
operators generating the asymptotic states. This fact is not totally
understood. The analysis of more complex theories in the formalism
proposed in this article might give new insights to this problem.

\resection{The models $M_{2,2p+3}$ \label{sec-2n}}

{}From our point of view, the models $\M_{2,2p+3}$ constitute the
simplest systems containing more than one asymptotic particle, since
their Kac table contains only one row of fields. Also their spectrum
contains only scalar particles, and from the analysis of the previous
sections we expect only scalar operators to appear in the massive
model, as it was also the case for the Yang-Lee model.

The $S$-matrices are given by \cite{fkm} \be S_{ab}=f_{\mid
  a-b\mid\frac\alpha 2}(\beta) f_{(a+b)\frac \alpha 2}(\beta)
\prod_{k=1}^{min(a,b)-1} (f_{(\mid a-b\mid +2k)\frac\alpha
  2}(\beta))^2 \sb , \label{sss} \ee where $\alpha=\frac
{2\pi}{2p+1}$, and $a,b=1,2,\ldots p$ label the particles of mass
$m_a=\sin\left (a \frac{\alpha}{2}\right )$.  The functions $f$ are
given by \be f_\alpha(\beta) \equiv \frac{\tanh \frac
  12(\beta+i\alpha)}{\tanh \frac 12 (\beta-i\alpha)} \sb .\ee
Note that the $S$-matrix of the fundamental particle has the same
functional structure as the $S$-matrix of the Yang-Lee model, the
difference being that the parameter $\alpha$ is specified differently.
We can therefore use  the results from section \ref{sec-yl}, where we had
analysed the kinematical recursion relation for the form-factors
corresponding to this $S$-matrix element for general $\alpha$.

The form factor equations for the models $\M_{2,2p+3} +\phi_{1,3}$
have been analysed in detail in \cite{a11}. There the form factors for
all perturbed primary operators $\phi_{1,k}$, $k=1\dots,p$ have been calculated
explicitly. They are given by (\ref{parafyl}) with
$Q_n = \Q_n(k)$ where $\Q_n(k)$ is defined in
(\ref{sol}).
Further it was shown  that there is only one
independent recursion relation resulting from all possible bound state
fusion processes.  This recursion relation can be expressed as a
constraint on the form factors with only indices $1\dots 1$ relating
$F_{n+2p+1}$ to $F_n$.  It follows that the consistent form factors of
this model form a subset of the solutions of the kinematical recursion
relation on particle 1. This allows us to determine the structure of
the form factors of higher particles by carrying out the bootstrap on
the expression obtained for $F_{1,1,\dots,1}$.

We carry out the bootstrap explicitly for the form factors involving
particles one and two.  They are given by \be
\F_{\underbrace{2,2,\dots,2}_q\underbrace{1,1,\dots,1}_r}(
\beta_1,\dots,\beta_{q+r} ) = \ee
$$H_{2q+r} Q_{2,2\dots2,1,1\dots1} \prod_{m=1}^q A(x_m) \prod _{i<j}^r
B_{1,1} (\beta_{ij} ) \prod_{m=1}^q\prod_{i=1}^r B_{1,2}
(\beta_{mi})\prod_{m<l}^q B_{2,2} (\beta_{ml}) \sb .$$ The
single components in this complex expression are  \be B_{1,1} (\beta) =
\frac{(-i) \sinh(\frac \beta 2) \zeta_{11}(\beta)} {(x_i+x_j) <\alpha>
  }\label{B11} \sb ,\ee \be B_{1,2} (\beta) = \frac{ -\zeta_{12}
  (\beta)}{4 {x_m x_i} \zeta(i \pi,\alpha )^2 <1-\frac \alpha
  2><\frac 32 \alpha> }
\label{B12}\sb ,\ee
\be B_{2,2} (\beta) = \frac{(-i) \sinh\frac \theta 2 \zeta_{22}
  (\beta)}{4 x_m x_n \pi \zeta(i \pi,\alpha)^4 <2
  \alpha><\alpha><1-\alpha> (x_m +x_n)^2}\label{B22}\sb .\ee $H_n$ is
the normalisation constant of $ F_{1,1,\dots,1}$ and \be A(x) = -
\cot \alpha \tan \frac \alpha 2 \frac 1 {\sqrt{\pi \zeta(i
    \pi,2 \alpha)}\, \zeta(i \pi,\alpha)\,2 \pi} \sb . \ee Further we have
introduced an abbreviated expression for the functions introducing the
bound state poles, \be <\alpha> \equiv \sinh \frac 12 ( \beta_{ij} - i
\alpha ) \sinh \frac 12 ( \beta_{ij} + i \alpha ) \sb ,\ee and the
minimal 2 particle form factors \be\ba{l} \zeta_{11} (\beta) =
\zeta(\theta,\alpha) \sb ,\\ \zeta_{12} (\beta) = \zeta(\theta,\frac
\alpha 2)\zeta( \theta,\frac {3\alpha}2) \sb ,\\ \zeta_{22} (\beta) =
\zeta(\theta,2\alpha) \zeta(\theta,\alpha)^2\sb ,\ea\ee

with $$ \zeta (\beta,\alpha) =\prod_{k=0}^\infty \frac{ \Gamma\left
  (k+\frac 12 +\frac \alpha {2\pi} -\frac{i \beta}2 \right )
  \Gamma\left (k+1 -\frac \alpha {2\pi} -\frac{i \beta}2 \right )}{
  \Gamma\left (k+\frac 12 -\frac \alpha{ 2\pi} -\frac{i \beta}2 \right
  ) \Gamma\left (k+\frac \alpha{ 2\pi} -\frac{i \beta}2 \right )}
\frac {\Gamma\left (k+\frac 32+ \frac \alpha {2\pi} +\frac{i \beta}2
\right ) \Gamma\left (k+2- \frac \alpha {2\pi} +\frac{i \beta}2 \right
) }{\Gamma\left (k+\frac 32 -\frac \alpha {2\pi} +\frac{i \beta}2
\right ) \Gamma\left (k+1+ \frac \alpha {2\pi} +\frac{i \beta}2 \right
) }\, .$$
We start
our investigation by analysing the system $\M_{2,7}$.
The above expressions are sufficient to examine the operator content of
this model, since it contains only two particles.  Let us analyse the form
factors corresponding to chiral operators, {\em i.e.} $N=0$ in
(\ref{ansatz}). This implies that the functions $Q_{2,2\dots
  2,1,1\dots1}$ are polynomials.

As before, we apply the counting technique, but now we have to take in
account that we treat a two-particle system.  The recursion problem
for the functions $Q_{\e_1,\dots\e_n}$ is now of  matrix-form.
We need to determine the respective degrees of the polynomials
$Q_{\e_1,\dots\e_n}$ and of the kernel of the recursion relation. The
degrees of $Q_{2,2\dots2, 1,1\dots 1}$ follow from Lorentz invariance,
as
\be deg(Q_{\underbrace{2,2,\dots,2}_q,\underbrace{1,1,\dots,1}_r})=\frac
12 (2 q+r) (2 q+r -1)-q+s\sb ,\label{degq}\ee where $s$ denotes the spin of the
operator as usual.  The degrees of the kernels of the recursion
relations, $\K_{2,2,\dots,2,1,1,\dots,1}$ follow  from the pole
structure of (\ref{B11}) -(\ref{B22}), and are given by
\be deg (\K_{\underbrace{2,2,\dots,2}_q,\underbrace{1,1,\dots,1}_r}) =
\frac 32 r (r-1) + 4 r q + 4 q(q-1) \sb .\label{degk}\ee

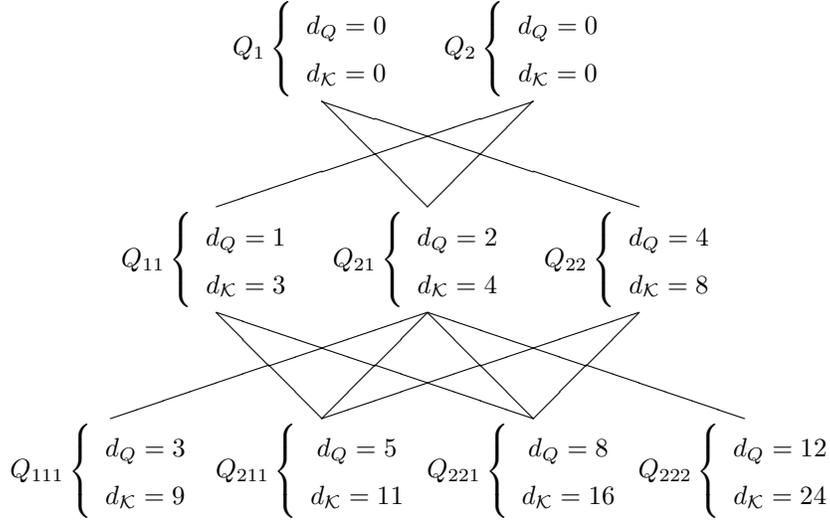
\begin{figure}
\begin{picture}(400,180)(-200,-180)
\put(-40,0){\makebox(0,0){\footnotesize $Q_1 \left \{ \ba{c} d_Q  = 0 \\
                                               d_\K = 0 \ea
                               \right .$ }}
\put(40,0){\makebox(0,0){\footnotesize $Q_2 \left \{ \ba{c} d_Q  = 0 \\
                                               d_\K = 0 \ea
                               \right .$ }}
\put(-80,-80){\makebox(0,0){\footnotesize $Q_{11} \left \{ \ba{c} d_Q  = 1 \\
                                               d_\K = 3 \ea
                               \right .$ }}

\put(0,-80){\makebox(0,0){\footnotesize $Q_{21} \left \{ \ba{c} d_Q  = 2 \\
                                               d_\K = 4 \ea
                               \right .$ }}
\put(80,-80){\makebox(0,0){\footnotesize $Q_{22} \left \{ \ba{c} d_Q  = 4 \\
                                               d_\K = 8 \ea
                               \right .$ }}
\put(-120,-160){\makebox(0,0){\footnotesize $Q_{111}\left\{ \ba{c} d_Q  = 3 \\
                                               d_\K = 9 \ea
                               \right .$ }}
\put(-40,-160){\makebox(0,0){\footnotesize $Q_{211} \left \{ \ba{c} d_Q  = 5 \\
                                               d_\K = 11 \ea
                               \right .$ }}
\put(40,-160){\makebox(0,0){\footnotesize $Q_{221} \left \{ \ba{c} d_Q  = 8 \\
                                               d_\K = 16 \ea
                               \right .$ }}
\put(120,-160){\makebox(0,0){\footnotesize $Q_{222} \left \{ \ba{c} d_Q  =12 \\
                                               d_\K = 24 \ea
                               \right .$ }}
\put(0,-60){\line(1,1){40}}
\put(0,-60){\line(-1,1){40}}
\put(80,-60){\line(-3,1){120}}
\put(-80,-60){\line(3,1){120}}
\put(40,-140){\line(1,1){40}}
\put(40,-140){\line(-1,1){40}}
\put(-40,-140){\line(1,1){40}}
\put(-40,-140){\line(-1,1){40}}
\put(40,-140){\line(-3,1){120}}
\put(-120,-140){\line(3,1){120}}
\put(120,-140){\line(-3,1){120}}
\put(-40,-140){\line(3,1){120}}
\end{picture}
\caption{Comparing dimensions of $Q_{\e_1,\dots,\e_n}$ with
    the respective kernels of the recursion relations. The lines
    indicate the inter-relations due to the bound state equation
    (\protect\ref{bounds}).}
\label{fig-27}
\end{figure}

We have depicted these dimensions in figure \ref{fig-27}. As in the
one particle systems we use this comparison of degrees in order to
count the number of independent solutions of the recursion relations.
Note that the polynomials $Q_{\e_1,\dots\e_n}$ are symmetric in
coinciding indices only.  Therefore in this example we have two sets
of indices, which have to be taken in account separately.  From figure
\ref{fig-27} we immediately find that the first solutions are being
generated by \be \frac 1{(q)_1} + \frac 1 {(q)_1} \sb ,\ee which
correspond to the possible choices of one-particle form factors.
The terms \be
\frac{q^2}{(q)_2}+ \frac{q^2}{(q)_1 (q)_1}+ \frac{q^4}{(q)_2} \sb ,\ee
generate the number of independent solutions for the two-particle form
factors.  The power of $q$ in the nominator derives from the
spin-value at which $deg(Q_{\e_1,\e_2}) +s =
deg(\K_{\e_1,\e_2})$. In particular, the
denominator $(q)_1 (q)_1$ corresponds to $Q_{21}(x_1,x_2)$ and indicates that
this function is not symmetric with respect to its two arguments.
Finally for the three-particle form factors we find \be
\frac{q^6}{(q)_3} + \frac{q^6}{(q)_2 (q)_1}+ \frac{q^8}{(q)_1 (q)_2}+
\frac{q^{12}}{(q)_3} \sb .\ee

In order to extract the general generating function, one uses the
fact, that the spin value at which a new solution enters is given by
$deg(\K_{\e_1,\dots\e_n}) -deg(Q_{\e_1,\dots\e_n})$.  Defining \be
F_{a,b} \equiv \sum_{n_1,n_2} \frac{q^{(n_1+n_2)^2 +n_2^2-a (n_1+n_2)
    - b n_2}}{(q)_{n_1} (q)_{n_2}}\sb , \label{fab}\ee and using
(\ref{degq}) and (\ref{degk}) one finds that the space of chiral
operators in the $\phi_{1,3}$ perturbation of $\M_{2,7}$ is generated
by $F_{1,1}$.  The functions (\ref{fab}) satisfy \bea F_{a,b} &=&
F_{a,b-1} + \q^{a+b-2} F_{a-2,b-2} \sb ,\nonumber\\ F_{a,b} &=&
F_{a-1,b+1} + \q^{a-1} F_{a-2,b} \sb .\label{rec27}\eea These
recursion relations can be used to prove that \be F_{1,1} = F_{-1,-1}
+ F_{0,-1} + F_{0,0} = \c_{1,1}(q) + \c_{1,2}(q) +\c_{1,3}(q) \sb .\ee
As in the previous cases we find that the chiral operator content of
the massive model is in a one-to-one correspondence with that of the
conformal model.

Unfortunately the results for this model are less complete than the
others. We have tried to solve the recursion relations (\ref{rec27})
to find an exact expression for the whole space of operators (given by
$\lim_{N\to \infty} F_{N,N}$) of the model, but did not succeed.
Nevertheless we have solved them numerically up to order $50$, and
the results indicate that \be F_{N,N} = \c_{1,1}(q) \c_{1,1} (\q) +
\c_{1,2}(q) \c_{1,2}(\q) +\c_{1,3}(q) \c_{1,3}(\q) \sa (mod \,q^N) \sb
, \ee
as expected.

Finally let us turn to the general systems $\M_{2,2p+3}$.  Again we
need to determine the dimensions of $Q_{\e_1,\dots,\e_n}$ and the
kernel $\K_{\e_1,\dots,\e_n}$.  Denote as $Q^{m_1,m_2,\dots,m_p}$ the
function $Q_{\e_1,\dots,\e_n}$ corresponding to $m_i$ indices $\e=i$,
where $i=1,\dots,p$, the number of different asymptotic states in the
theory.  From Lorentz invariance one finds that
$$ deg(Q^{m_1,m_2,\dots, m_p}) = \frac 12 (\sum_{i=1}^p i m_i)
(1+\sum_{i=1}^p i m_i) - \sum_{i=1}^p \frac 12 i(i-1) m_i\sb .$$ For
example the dimension of $Q_{kl}(x_1,x_2)$, i.e. the function corresponding to
the two-particle form factor of particles $k$ and $l$, has dimension
$deg(Q_{kl}) =kl$\,.

In order to determine the dimension of the kernel we must carry out
the bootstrap on the form factors of particle 1, satisfying the
kinematical recursion relation. This is straightforward for the
denominators and for the factor $\sinh (\frac\beta 2)$ in (\ref{B11}).
But one needs to be careful, since further zeros appear from the
bootstrap of the factor $\zeta_{11}$. For example for the two particle
form factor $\F_{kl}$ with $k >l$ one finds through the bootstrap that
$$ \zeta_{1,1} \to \left (\sinh \frac \beta 2\right )^{\delta_{n,k}}
\zeta\left (\beta,(k-l)\frac \alpha 2\right )\, \zeta\left
(\beta,(k+l)\frac \alpha 2\right )\, \prod_{i=1}^{l-1} \zeta\left
(\beta,(k-l+2i)\frac \alpha 2 \right )\, \times
$$ \be \prod_{i=2}^{k-1} \prod_{j=0}^{l-1} \zeta\left
(\beta,(k+l)\frac \alpha 2 -(i+j) \alpha \right ) \sb .\label{zeros}
\ee The first line in (\ref{zeros}) gives exactly the minimal
two-particle form factor for the particles $k$ and $l$,$\zeta_{kl}$,
 while the
second line reduces to trigonometric functions, due to the relation
$\zeta(\beta,\alpha) \zeta(\beta,-\alpha) = <\alpha >$.  They cancel
poles which arise from the bootstrap of the poles in the denominator
of (\ref{B11}).

We will not give a detailed account of this
calculation. Carrying out the bootstrap on all factors in
(\ref{parafyl}) one finds that the total degree of the kernel of the
recursion relations corresponding to the function $Q_{kl}$ is given by
$deg(\K_{kl}) = k.l + 2.l$.  Comparing with the dimensions of
$Q_{kl}$, one finds that solutions corresponding to the kernel of
$\K_{kl}$ will enter at level $2 l$ (recall we have the convention
that $k>l$).  Therefore the general exponent in the $q$-sum expression
will be given by
$$ \ba{l} 2 l\, n_k n_l \sb {\rm for} \,\, k > l \sb ,\\ k \, n_k (n_k
-1) \sb {\rm for}\,\, k =l \sb ,\ea $$ for every pair of indices $n_k$
and $n_l$. Taking the sum over all particles we find \be \sum_{k=1}^p
\left ( k n_k (n_k-1) + \sum_{l<k} 2 l n_l n_k \right ) =\sum_{k=1}^p
  N_k^2-N_k \sb ,\ee where we defined $N_j = \sum_{i=0}^{p-j}
  n_{p-i}$. This leads to the final expression for chiral massive
  states as \be F_{1,1,\dots,1}=\sum_{n_1,\dots n_p} \frac{
    q^{N_1^2-N_1 + N_2^2 -N_2 + \dots + N_p^2-N_p}} {(q)_{n_1}
    (q)_{n_2} \dots (q)_{n_p}} \sb .\ee

  We want to show that this sum expression coincides with a sum over
  characters of the model. For that we need the $q$-sum expressions of
  the characters, given through the Gordon-Andrews identities (see
  {\em e.g.} \cite{andrews}) as \be \c_{1,j}(q) =
  F_{\underbrace{00\dots 0}_{j-1}\underbrace{-1-1\dots -1}_{p+1-j}}
  \sb , \ee where we have introduced the generating functions \be
  F_{k_1,\dots ,k_p} \equiv \sum_{n_1,\dots n_p} \frac{ q^{N_1^2-k_1
      N_1 + N_2^2 -k_2 N_2 + \dots + N_p^2- k_p N_p}} {(q)_{n_1}
    (q)_{n_2} \dots (q)_{n_p}} \sb ,\ee which satisfy the recursion
  identities $$ F_{k_1, \dots, k_p} = F_{k_1,\dots, k_{i-1} ,
    k_i-1,k_{i+1}+1, k_{i+2},\dots,k_p} + $$ \be
  \q^{k_1-1}\q^{k_2-1}\dots\q^{k_i-1}
  F_{k_1-2,k_2-2,\dots,k_i-2,k_{i+1}, \dots, k_p} \sb .\label{rec2n}
  \ee

Apply these identities to the generating function of the chiral
operators, namely
$$ F_{11\dots 1} = F_{111\dots 10} + F_{-1-1\dots-1-1} \sb .$$ The
last factor corresponds to $\c_{1,1}(q)$. Now we have to continue with
$$F_{11\dots 110} = F_{11\dots 01} + F_{-1-1,\dots -10}\sb .$$ This
last term does not correspond to any conformal character, but if we
use the recursion relations to permute through the $0$ to stand at the
first position we find $\c_{1,2}(q)$ along with a plethora of terms,
without physical interpretation. If one carries out this commutation
process of generating always further zeros as the last index and
commuting them through, one will finally end up with \be F_{11\dots1}
= \sum_{i=1}^{p} \c_{1,i}^{(2,2p+3)}(q) \sb .\ee

We conclude that in this series of models the operator content of the
massive theory coincides with the content of conformal scalar
operators.  Unfortunately in all these systems, containing more than
one asymptotic state, we were not able to solve the general
recursion-relations (\ref{rec2n}) which determine the
full space of operators of the massive model. We have done several
numerical calculations which lead us to the conjecture that the full
space of operators in the massive model corresponds to the scalar
conformal one, {\em i.e.} \be Z = \sum_{i=1}^{p} \c_{1,i}(q)
\c_{1,i}(\q) \sb .\ee

We have also compared the structure of the solutions of these
recursion relation with several known `finitized' expressions for the
characters of these models \cite{quano}, but were not able to find one
which provides a solution to the recursion problem (\ref{rec2n}).

\resection{Conclusions}

The aim of this work was to analyse the structure of the space of
operators in massive integrable systems. Our particular interest was
in deformed minimal conformal field theories, because these theories
at the critical point have a well-known structure, that is the
operators are organised into families determined through the Virasoro
irreducible representations, with a highest weight state corresponding
to a primary operator with rational conformal weight.  All other
operators can be obtained by applying the generators of the Virasoro
algebra to these primary fields.

For massive integrable models such an algebraic analysis has not yet
been developed.  As an alternative we have used the form factor
bootstrap method in order to determine the structure of the space of
operators. In order to find similarities we have used the physical
properties of the conformal operators and have searched for analogues
in the massive models.  In particular we have introduced a grading into
our space of operators in the massive model defining chiral operators
as those with the mildest ultraviolet behaviour and through their
spin. The whole space of operators is constructed by augmenting the
divergence of the form factors gradually.

Using the recursive equations for the form factors we have proposed a
counting procedure in order to find the number of operators in each of
the finite dimensional subspaces defined through this grading.  Though
we have analysed only simple systems (in the sense that the Kac-table
of the corresponding conformal model consisted at the most of two
rows), the method is applicable to {\em any} system described through
a factorised scattering theory, including systems with degenerate
particles and even  massless scattering theories, which describe the
flow from one critical point to another.

Our analysis of the massive models gives several interesting results.
For all the systems examined the chiral field content is in a
one-to-one correspondence with that of the underlying conformal
theory. The massive primary operators are grouped into families
according to the parity of the corresponding conformal field.  This
induces the possibility of a mixing of the descendent operators in the
corresponding Virasoro irreducible representations in the process of
perturbation.  The full space of space of operators in the massive
model contains two types of fields. For theories with scalar
asymptotic states there are only scalar fields, while for models with
asymptotic particles which satisfy non-trivial statistics, in addition
to the scalar fields also para-fermionic operators are present in the
theory.  This generates for example in the model $\M_{3,5}$ operators
satisfying para-statistics with fractional spin $\frac 14$.

Finally let us comment on the mathematical methods involved.  A
feature, which persisted in all models examined, was that the counting
of the operators lead naturally to fermionic sum expressions of the
characters, or rather sums of them. Recently a lot of interest has
evolved in this subject, after Kedem {\em et.al.} \cite{kedem} have
conjectured a whole series of such character identities, which have
been proved in \cite{quano,melzer,berkovich}, by using finitizations
of the character expressions. As we have seen, the counting procedure
leads to finite expressions for the characters, when we generate the
full space of states. It seems to be an interesting fact, that the
method here proposed gives a natural physical application to these
mathematical results.

\newpage \subsection*{Acknowledgments} I am indebted to A. Berkovich,
L. Chau, A. Fring, R. Kedem, A. Kent, B. McCoy, G. Watts and A.
Zamolodchikov for discussions, and to A. Recknagel for obtaining some
important literature for me. This work was supported by PPARC grant
GR/J20661.

\end{document}